\documentclass[fleqn,usenatbib]{mnras}

\usepackage{newtxtext,newtxmath}

\usepackage[T1]{fontenc}

\DeclareRobustCommand{\VAN}[3]{#2}
\let\VANthebibliography\thebibliography
\def\thebibliography{\DeclareRobustCommand{\VAN}[3]{##3}\VANthebibliography}

\usepackage{graphicx}	
\usepackage{amsmath}	

\usepackage{multirow} 
\usepackage{threeparttable}

\newcommand{\ha}{\hbox{H$\alpha$}}

\newcommand{\hii}{\hbox{H\,{\sc ii}}}
\newcommand{\oiii}{\hbox{[O\,{\sc iii}]}}

\newcommand \Lsun{L_\odot}

\def\gtsim{~\rlap{$>$}{\lower 1.0ex\hbox{$\sim$}}}
\def\ltsim{~\rlap{$<$}{\lower 1.0ex\hbox{$\sim$}}}

\title[SED of NGC~1365]{Radio-to-Submillimetre Spectral Energy Distributions of NGC~1365} 

\author[Chen et al.]{
Guangwen Chen$^{1,2,3}$\thanks{E-mail: guangwen@mail.ustc.edu.cn}, 
George J. Bendo$^{4}$, 
Gary A. Fuller$^{4,5}$,
Hong-Xin Zhang$^{1,2}$, 
and 
Xu Kong$^{1,2,6}$
\\
$^{1}$Department of Astronomy, University of Science and Technology of China, Hefei 230026, China\\
$^{2}$School of Astronomy and Space Sciences, University of Science and Technology of China, Hefei 230026, China\\
$^{3}$Jodrell Bank Centre for Astrophysics, Department of Physics and Astronomy, The University of Manchester, Oxford Road, Manchester M13 9PL, United Kingdom \\
$^{4}$UK ALMA Regional Centre Node, Jodrell Bank Centre for Astrophysics, Department of Physics and Astronomy, The University of Manchester,\\ Oxford Road, Manchester M13 9PL, United Kingdom\\
$^{5}$I. Physikalisches Institut, University of Cologne, Z\"ulpicher Str. 77, 50937 K\"oln, Germany\\
$^{6}$Frontiers Science Center for Planetary Exploration and Emerging Technologies, University of Science and Technology of China, Hefei, Anhui, 230026, China \\
}

\date{Accepted 2024 March 21. Received 2024 March 18; in original form 2024 January 24}

\pubyear{2024}

\begin{document}
\label{firstpage}
\pagerange{\pageref{firstpage}--\pageref{lastpage}}
\maketitle

\begin{abstract}

We analyse the radio-to-submillimetre spectral energy distribution (SED) for the central pseudobulge of NGC~1365 using archival data from the Atacama Large Millimeter/submillimeter Array (ALMA) and the Very Large Array (VLA). 
This analysis shows that free-free emission dominates the continuum emission at 50--120~GHz and produces about 75 per cent of the 103~GHz continuum emission. 
However, the fraction of 103~GHz continuum emission originating from free-free emission varies significantly among different subregions in the pseudobulge, particularly for an outflow from the AGN on the eastern pseudobulge where the synchrotron emission produces half of the 103~GHz continuum emission. 
Free-free emission also dominates at 103~GHz within the central 400 pc diameter region, but this emission is associated with the AGN rather than star formation.
The star formation rate (SFR) within the pseudobulge derived from the ALMA free-free emission is $8.9 \pm 1.1$~M$_\odot$~yr$^{-1}$.   
This is comparable to the SFR from the mid-infrared emission but higher than the SFR from the extinction-corrected $\ha$ line emission, mainly because the pseudobulge is heavily dust obscured.  
The 1.5 GHz emission yields a comparable SFR for the pseudobulge but may have lower SFRs within subregions of the pseudobulge because of the diffusion outside of these regions of the electrons producing the synchrotron radiation.
We propose that applying a correction factor of 75 per cent to the 80--110~GHz continuum emission could provide valuable estimates of the free-free emission without performing any SED decomposition, which could derive extinction-free SFRs within 20 per cent accuracy.

\end{abstract}

\begin{keywords}
galaxies: individual: NGC 1365 -- galaxies: starburst -- galaxies: star formation -- radio continuum: galaxies.
\end{keywords}



\section{Introduction}\label{sec:intro}

The spectral energy distributions (SEDs) of galaxies from the radio to the submillimetre include contributions from synchrotron emission, free-free emission and thermal dust emission (as well as potentially more exotic forms of emission such as anomalous microwave emission).  Synchrotron emission mainly originates from active galactic nuclei (AGNs) and/or supernova remnants, free-free emission arises from the gas surrounding photoionizing stars, and thermal dust emission emanates from interstellar dust \citep{Condon1992}.  All of these forms of emission within this frequency range are generally unaffected by dust obscuration.  Therefore, the decomposition of the SEDs can provide a direct approach to understanding the AGNs and star formation within galaxies.

However, the analysis of SEDs from the radio to the submillimetre continuum emission have only been published for a small number of galaxies using relatively heterogeneous data, including M31 \citep{Planck2015,Battistelli2019,Harper2023,Fernandez2024}, M82 \citep{Condon1992,Peel2011}, NGC~253 \citep{Peel2011,Bendo2015}, NGC~1808 \citep{Salak2017,Chen2023}, NGC~3256 \citep{Michiyama2020} and NGC~4945 \citep{Peel2011,Bendo2016}.   
Only \citet{Michiyama2020} and \citet{Chen2023} discussed variations of the fractions of the total millimetre continuum emission from synchrotron emission, free-free emission and thermal dust emission within different regions in the individual galaxies that they studied.  \citet{Linden2019} and \citet{Linden2020} presented analyses of spectral indices for a larger sample of local luminous infrared galaxies (LIRGs; $L_{\rm IR}>10^{11}\Lsun$) based on 3, 15 and 33~GHz continuum emission and revealed the connection between synchrotron radiation and free-free emission within this frequency range for their galactic nuclei and for extranuclear star-forming regions.  However, their data could not fully constrain the shape of the SEDs and the proportions of thermal dust emission, free-free emission and synchrotron emission in the submillimetre and millimetre bands.  
Clearly, the scientific literature still lacks detailed information on the radio-to-submillimetre SEDs of galaxies in spite of the importance of these bands in terms of measuring star formation rates and studying AGN activity.

In this paper, we report an analysis of the 2--246~GHz SED for the central pseudobulge of NGC~1365 based on data from the Atacama Large Millimeter/submillimeter Array (ALMA) and the Very Large Array (VLA).   This object is a barred spiral galaxy located at a distance of $19.6\pm0.8$ Mpc \citep[][]{Anand2021a} within the Fornax Cluster \citep{Lindblad1999}.  It hosts a Seyfert 1.5 - 2 nucleus \citep[e.g.,][]{Turner1993,Maiolino1995,Hjelm1996,Sakamoto2007} and, the AGN is well detected in X-ray observations \citep[e.g.,][]{Risaliti2005,Wang2009,Swain2023b}.  This galaxy has intense star formation and a biconical outflow in the central region, which is heavily dust obscured \citep[e.g.,][]{Venturi2017,Fazeli2019,Gao2021}. 
The SEDs from the far-infrared to the near-infrared in the centre at a resolution of 20 arcsec has been presented by \citet{Tabatabaei2013}, and the SEDs from the infrared to the X-ray for the nucleus has been presented by \citet{Swain2023b}, but the radio-to-submillimetre SEDs are still unknown.  The galaxy has been identified as a LIRG \citep[e.g.,][]{Armus2009}, with most of the infrared emission originating from the central pseudobulge.  The high surface brightness of this compact region across multiple bands makes it an ideal object for submillimetre and millimetre observations.  In addition, ancillary data from the from the {\it Spitzer} Space Telescope \citep{Werner2004} and the {\it Herschel} Space Observatory \citep{Pilbratt2010} and recent James Webb Space Telescope (JWST) observations \citep[][]{Whitmore2023,Schinnerer2023,Liu2023} provide supplemental infrared data for this object.

We present the description of data in Section \ref{sec:data}.  Continuum images for the centre of NGC~1365 spanning radio to submillimetre wavelengths are presented in Section \ref{sec:images}.  The analysis of the SEDs is described in Section \ref{sec:SED}, while Section \ref{sec:SFR} presents the measurements of star formation rates (SFRs) and comparisons of SFRs from the free-free emission with the SFRs from the radio continuum emission and infrared dust emission.  A brief summary of our main results is given in Section \ref{sec:conclusion}.

\section{Data}\label{sec:data}

To accurately characterise the SEDs spanning from 2 to 246~GHz, we use the National Radio Astronomy Observatory (NRAO) VLA L-, S-, C-, and X-band data (covering 2--10~GHz) and the ALMA Band 3 and 6 data (covering 100--246~GHz).  
These data had $uv$ ranges that overlapped significantly (in terms of $k\lambda$) that could therefore be used to create SEDs with comparable beam sizes and maximum recoverable scales.  The ALMA and VLA archives as well as the archive for the Australia Telescope Compact Array (ATCA) contained additional data covering other frequencies, but these data differed from the data listed above in terms of $uv$ coverage and thus yield images with either beam sizes that are too large or maximum recoverable scales that are too small for the analysis in the main part of this paper.  We instead used these data in supplemental SED analyses in Appendices \ref{appendix:dust} and \ref{appendix:ame}.  24~$\mu$m data from the {\it Spitzer} Space Telescope are used to calculate alternate SFRs for comparisons. 

\subsection{ALMA data} \label{subsec:ALMA}   

Archival ALMA data\footnote{\url{https://almascience.eso.org/aq/}} from multiple projects are used in our analysis; see Table~\ref{tab:ALMA_obs} for details.  We did not use the Band 7 and 8 data in constructing the SEDs that are analyzed in Section \ref{sec:SED} because their maximum recoverable scales are not large enough to cover the whole circumnuclear region within the galaxy. 
As seen in that section, the thermal dust emission is the dominant source of continuum emission in ALMA Band 6.   However, in Appendix \ref{appendix:dust}, we examined how including Band 7 and 8 data affected the SED fitting.  ALMA project 2019.2.00134.S also acquired Band 3 data, but we did not use these data because the beam size of the data is much larger in comparison to the other archival data in our analysis.   In addition, we did not include any ALMA total power observations in our analysis because those data cannot be used for continuum imaging.

\setlength{\tabcolsep}{5pt}

\begin{table*}
\centering
\begin{minipage}{1.0\textwidth}
\caption{General information of ALMA observations$^a$.} 
\label{tab:ALMA_obs}
\begin{tabular}{@{}ccccccccccc@{}}
    \hline
    \hline
      Band &
      Project &
      Array &
      Scheduling &
      5\%$-$80\% of  &
      Angular &
      Maximum &
      Frequency & 
      {\sc casa} \\
       &
      code &
       &
      Block &
      baselines$^b$ &
      resolution & 
      recoverable scale & 
      range & 
      version for \\
       &
       &
       &
       &
      (m) &
      (arcsec) &
      (arcsec) &
      (GHz) &
      calibration$^c$ \\
    \hline
      3 &
      2015.1.01135.S &
      12 m &
      NGC1365\_a\_03\_TE &
      43$-$225 &
      1.4 &
      13.4 &
      98.97-102.94, 111.52-114.88 & 
      4.7.0 \\
       &
      2017.1.00129.S &
      7 m &
      ESO358-G\_a\_03\_7M &
      9$-$32 &
      10.0 &
      77.8 &
      99.90-103.76, 111.84-115.70 & 
      5.4.0 \\
    \hline
      6 &
      2013.1.01161.S &
      12 m &
      NGC1365\_a\_06\_TC &
      25$-$144 &
      1.0 &
      11.5 &
      228.46-231.90, 242.67-246.22 & 
      4.3.1 \\
       &
       &
       &
      NGC1365\_a\_06\_TE &
      44$-$241  &
      0.6 &
      6.0 &
      228.46-231.90, 242.67-246.22 & 
      4.3.1 \\
       &
       &
       &
      NGC1365\_b\_06\_TE &
      134$-$785  &
      0.2 &
      1.9 &
      228.46-231.90, 242.67-246.22 & 
      4.3.1 \\
       &
       &
      7 m &
      NGC1365\_a\_06\_7M &
      7$-$27 &
      5.3 &
      36.6 &
      228.40-231.96, 242.61-246.28 & 
      4.3.1 \\
       &
      2021.1.01150.S &
      12 m &
      NGC\_1365\_a\_06\_TM1 &
      82$-$739 &
      0.2 &
      3.1 &
      225.79-230.23, 241.68-245.67 & 
      6.2.1 \\
       &
       &
       &
      NGC\_1365\_a\_06\_TM2 &
      24$-$152 &
      1.0 &
      10.8 &
      225.79-230.23, 241.68-245.67 & 
      6.2.1 \\
       &
      2021.1.01295.S &
      7 m &
      NGC\_1365\_a\_06\_7M &
       9$-$29 &
       4.9 &
       30.7 &
       226.27-230.30,
       240.70-244.67 & 
      6.2.1 \\
\hline
\end{tabular}
\\
$^a$This information is retrieved from ALMA Science Archive (\url{https://almascience.eso.org/aq/}). \\
$^b$These values are the 5th and 80th percentile of all projected baselines calculated from the $uv$ distributions of the observations, but not all data between the longest and shortest baseline are used in continuum imaging (see also Table \ref{tab:ALMA_continuum}).  \\
$^c$This is the specific version of {\sc casa} used to calibrate the data, which adheres to the general recommendations for pipeline-calibrated data.  {\sc casa} version 6.4.1 was used for imaging.
\end{minipage}
\end{table*}

\setlength{\tabcolsep}{6pt}

These ALMA visibility data were reprocessed using the {\sc common astronomy software applications} ({\sc casa}; \citealt{CASA2022}).  The general recommendation for recreating the pipeline-calibrated data is to use the same version of {\sc casa} that was originally used to calibrate the data; the specific versions that we used are listed in Table~\ref{tab:ALMA_obs}.   According to the ALMA Technical Handbook\footnote{\url{https://almascience.eso.org/documents-and-tools/cycle10/alma-technical-handbook}} \citep{Cortes2023}, the typical flux calibration uncertainties are 5 per cent for Band 3 and 10 per cent for Band 6 and 7 and 20 per cent for Band 8 data.

These processed visibility data were converted into continuum images using {\sc tclean} in {\sc casa} version 6.4.1.  We first created image cubes without any continuum subtraction to identify the line-free regions in the data.  After this, continuum images were created using visibility data from adjacent spectral windows with comparable frequencies.  When creating these images, we selected the H{\"o}gbom cleaning algorithm because it is effective for imaging data with relatively low signal-to-noise ratios.  For the Band 3 and 6 data, we used natural weighting to recover extended emission and to maximize the signal-to-noise ratio.  In addition, the $uv$ coverages were adjusted so that all images have the same full width at half-maximum (FWHM) of the restoring beam (2.1 arcsec) and the same maximum recoverable scale (36 arcsec). Data with relatively long $uv$ baselines (Band 3 data with baselines $>360$ m and Band 6 data with baselines $>145$ m) were not used in the continuum imaging.
Notably, the FWHM and maximum recoverable scale are not directly determined by the upper and lower limits of the $uv$ coverage itself but rather by the overall $uv$ distribution.  For instance, the Band 3 data have a $uv$ coverage with a large upper limit but fewer antennae on longer baselines, while the Band 6 data have a $uv$ coverage with a lower upper limit but more antennae on longer baselines; both dataset yield images that have restoring beams with similar FWHMs.  Consequently, the upper and lower limits of $uv$ coverages in units of kilolambda (k$\lambda$) may vary sightly across different frequency bands when creating images with comparable FWHMs and maximum recoverable scales. 
Primary beam corrections were applied to all images.  The pixel scales are set as 0.2 arcsec to subsample the beam effectively, and the image dimensions are set as $640\times640$ spaxels ($128\times128$ $\rm {arcsec^2}$ (corresponding to $\sim$12~kpc at a distance of 19.6~Mpc) to cover the primary beam.
Details related to the continuum imaging are listed in Table \ref{tab:ALMA_continuum}.

\begin{table}
\centering
\begin{minipage}{0.35\textwidth}
\caption{Information of ALMA continuum imaging.}
\label{tab:ALMA_continuum}
    \begin{tabular}{@{}cccc@{}}
    \hline
    \hline
      Band &
      Frequency &
      $uv$ range &
      Beam FWHM\\
       &
      (GHz) &
      (k$\lambda$) &
      ($\rm {arcsec^2}$)\\
    \hline
      3 &
      102.5 &
      $5.4-123$  &
      2.3 $\times$ 1.8\\%
       &
      113.5 &
      $5.4-131$ & 
      2.3 $\times$ 1.8\\%
    \hline
      6 &
      230.5 &
      $5.4-111$ & 
      2.2 $\times$ 1.9\\%
       &
      244.9 &
      $5.6-97$ & 
      2.2 $\times$ 1.9\\%
    \hline
    \end{tabular}\\
\end{minipage}
\end{table}

\subsection{VLA data}\label{subsec:VLA}  

The VLA visibility data from projects 17B-331, 18A-418, SH0126, and 21A-282 used for our SED analysis were downloaded from the NRAO archive\footnote{\url{https://data.nrao.edu}}.  We only used data with $uv$ ranges that allowed us to create images with beam FWHMs and maximum recoverable scales that are comparable to the ALMA continuum images.  For projects 17B-331 and 18A-418, we used the pipeline-calibrated data provided by the NRAO archive, and for projects SH0126 and 21A-282, we used {\sc casa} version 5.3.0 with the VLA scripted calibration pipeline 1.4.2\footnote{\url{https://science.nrao.edu/facilities/vla/data-processing/pipeline/scripted-pipeline}} to calibrate their visibility data.  
We also used the 1.5 GHz visibility data from the NRAO VLA Archive Survey \citep[NVAS\footnote{\url{http://www.vla.nrao.edu/astro/nvas/}};][]{Condon1998,Crossley2007,Crossley2008} to create a 1.5~GHz image that could be used to measure SFRs, but we did not include 1.5 GHz data in our SED analysis in Section \ref{sec:SED} for two reasons.  First, the beam is somewhat larger than the beam that we can produce using the data from the other bands, and this mismatch in beam size will affect the photometry.  Second, data below 2 GHz may cause the synchrotron emission to appear flatter because the cosmic rays producing the higher frequency emission are more affected by aging effect \citep{Davies2006}.
The NVAS processed the raw visibility data from 1976 to 2006 with the {\sc Astronomical Image Processing System} VLA pipeline to make the archive data easily accessible to the astronomical community.  General information for these observations is listed in Table \ref{tab:VLA_obs}.  

\begin{table}
\centering
\begin{minipage}{0.365\textwidth}
\caption{General information of VLA observation.}\label{tab:VLA_obs}
    \begin{tabular}{@{}cccc@{}}
    \hline
    \hline
      Band &
      Project ID &
      Date &
      Frequency range\\
       &
       &
       &
      (GHz) \\
    \hline
      L &
      AS133 &
      1983 Nov 06&
      1.46$-$1.51 
      \\
      &
      AC348 &
      1993 Feb 01 &
      1.46$-$1.66 
      \\
      S &
      21A-282 &
      2021 Aug 07 &
      2.0$-$3.9 \\
       &
       &
      2021 Dec 31 &
      2.0$-$3.9 \\
       &
       &
      2022 Apr 16 &
      2.0$-$3.9 \\
      C &
      17B-331 &
      2017 Nov 23 &
      4.0$-$7.9 \\
       &
      18A-418 &
      2018 Apr 16 &
      4.0$-$7.9 \\
      X &
      SH0126 &
      2021 Sep 30 &
      8.0$-$9.9 \\
    \hline
    \end{tabular}\\
\end{minipage}
\end{table}

We used {\sc clean} in {\sc casa} version 5.6.1 to image the 1.5~GHz data because they were not in a format that could be read by {\sc tclean}.  We used {\sc tclean} in {\sc casa} version 6.4.1 to image the other VLA data. 
When creating these images, we again used the H{\"o}gbom cleaning algorithm. We only used data with baselines with lengths (in terms of k$\lambda$) that were similar to what was used for the ALMA data so as to obtain images with similar beam sizes and maximum recoverable scales. Given the declination of this source relative to the location of the VLA, the image exhibits an elongated beam, appearing extended along the major axis and narrow along the minor axis.  To adjust the shape of beam to be comparable to ALMA Band 3 and 6 images, we first set the {\sc weighting} parameter to {\sc briggs} and the {\sc robust} parameter to 0.5 to improve the spatial resolution. Subsequently, we used {\sc imsmooth} in {\sc casa} to smooth the beams so that they were comparable in shape to the beams of ALMA images except for the 1.5~GHz data, which has a relatively large beam if we smooth it.  All the final VLA images have pixel sizes of 0.2 arcsec and dimensions of $640 \times 640$ spaxels ($128 \times 128$ arcsec$^2$) to match the ALMA images.  
Based on the information from the VLA Observational Status Summary 2023A\footnote{\url{https://science.nrao.edu/facilities/vla/docs/manuals/oss/performance/fdscale}} the flux scaling uncertainties are less than 10 per cent for the L-, C-, S-, and X-band data.  Detailed information related to the continuum imaging is listed in Table \ref{tab:VLA_continuum}.

\setlength{\tabcolsep}{5pt}

\begin{table}
\centering
\begin{minipage}{0.45\textwidth}
\caption{Information of VLA continuum imaging.}\label{tab:VLA_continuum}
    \begin{tabular}{@{}ccccccc@{}}
    \hline
    \hline
      Band &
      Frequency &
      $uv$ range &
      Beam FWHM & 
      Beam FWHM \\ 
        &
      (GHz) &
      (k$\lambda$) &
      before smoothing &  
      after smoothing \\  
       &
       &
       &
      ($\rm {arcsec^2}$)&
      ($\rm {arcsec^2}$)\\
    \hline
      L &
      1.5  &
      3.4$-$183 & 
      3.4 $\times$  1.4& 
      - \\
      S &
      2.6 &
      3.6$-$307 & 
      2.6 $\times$ 0.8 & 
      2.7 $\times$ 1.7 \\
       &
      3.6 &
      3.5$-$425 & 
      2.0 $\times$ 0.7& 
      2.5 $\times$ 1.8\\ 
      C &
      5.0 &
      3.4$-$250 & 
      2.1 $\times$ 0.7& 
      2.2 $\times$ 1.9 \\ 
       & 
      7.0 &
      3.4$-$187 & 
       1.8 $\times$ 0.8 & 
       2.2 $\times$ 1.9 \\ 
      X &
      9.0 &
      3.4$-$1054 & 
    2.3 $\times$ 0.6&  
    2.5 $\times$ 1.8\\ 
    \hline
    \end{tabular}\\
\end{minipage}
\end{table}

\setlength{\tabcolsep}{6pt}



\subsection{Infrared data}\label{subsec:infrared}

We used mid-infrared data as an alternate star formation tracer for comparison to those derived from our radio and millimetre data.  JWST 21~$\mu$m data are available for this galaxy \citep{Lee2023}, but those data are heavily saturated, so we used 24~$\mu$m data from the {\it Spitzer} Space Telescope instead.  The specific 24~$\mu$m image that we used was originally created and published by \citet{Bendo2020} using the MIPS data Analysis Tools version 3.10 \citep{Gordon2005} and additional data processing steps described by \citet{Bendo2012}.  The {\it Spitzer} image has the beam FWHM of 6 arcsec and pixel sizes of 1.5 arcsec, and the flux calibration uncertainty is 4 per cent \citep{Engelbracht2007}.


\section{Images} \label{sec:images}

\begin{figure}
  \centering
  \includegraphics[width=0.48\textwidth]{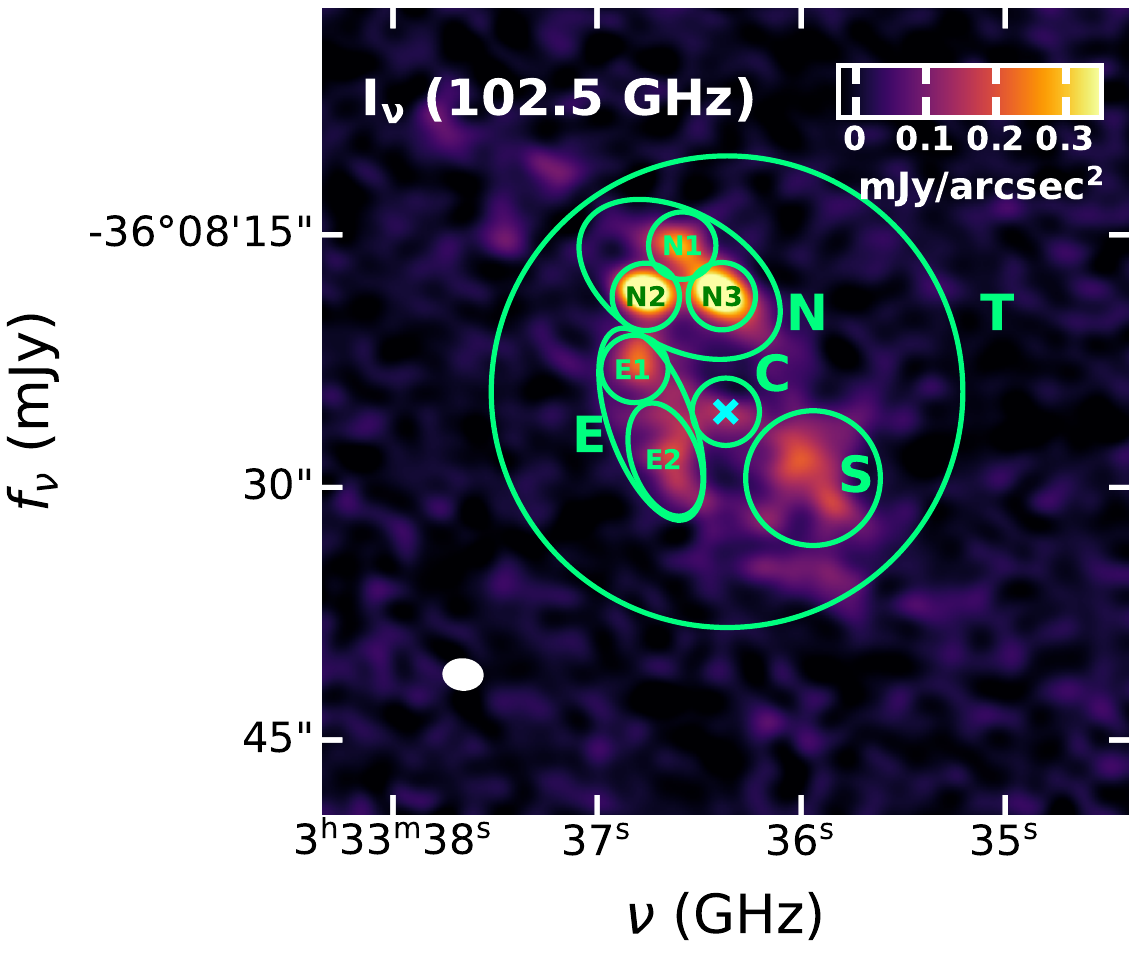} 
  \caption{The image of 102.5~GHz continuum emission in the central 48~arcsec of NGC~1365. The solid green circles identify the locations where we measured the continuum emission, including the total (T), north (N), east (E), centre (C) and south (S) with diameters of 28~arcsec (2.7~kpc), 13$\times$8~arcsec (1240$\times$760~pc), 12$\times$5~arcsec (1140$\times$480~pc), 4~arcsec (380~pc) and 8~arcsec (760~pc), respectively.
  Smaller apertures within the N, E and S region, including the N1, N2, N3 and E1 regions with diameters of 4~arcsec (380~pc) and the E2 region with 7$\times$4~arcsec (670$\times$380~pc), are also marked as solid green circles.  The cyan ``X'' symbol marks the location of the AGN, indicated by the peak of X-ray emission \citep{Wang2009}. The restoring beam is shown as the filled white ellipse in the lower-left corner.}
  \label{fig:Image_Aper}
\end{figure}

Figure \ref{fig:Image_Aper} shows the 102.5~GHz continuum emission in the central region of NGC~1365 and the locations where we measured continuum emission from radio to submillimetre bands for constructing SEDs.  We also refer to these regions when discussing the multiwavelength images qualitatively.   
The AGN is located within the centre (C) region.  The north (N) region represents the area where the gas flows in from the northeast along the bar, the south (S) region is the downstream location of the gas from the N region, and the east (E) region shows the downstream location where the gas flowing in from the southern arm. The N, S and E regions are comparable to the ``Northern Arm'', ``Mid-Southwest Region'' and ``Mid-East Region'' presented by \citet{Liu2023}.  In addition, the N region includes three bright sources, which we labelled as ``N1'', ``N2'' and ``N3'', and the E region includes two relatively faint sources, which we labelled as ``E1'' and ``E2''.  The total (T) aperture with a diameter of 28~arcsec (2.7~kpc) was used to measure the total detectable emission from the central region. 

\begin{figure*}
  \centering
  \includegraphics[width=1\textwidth]{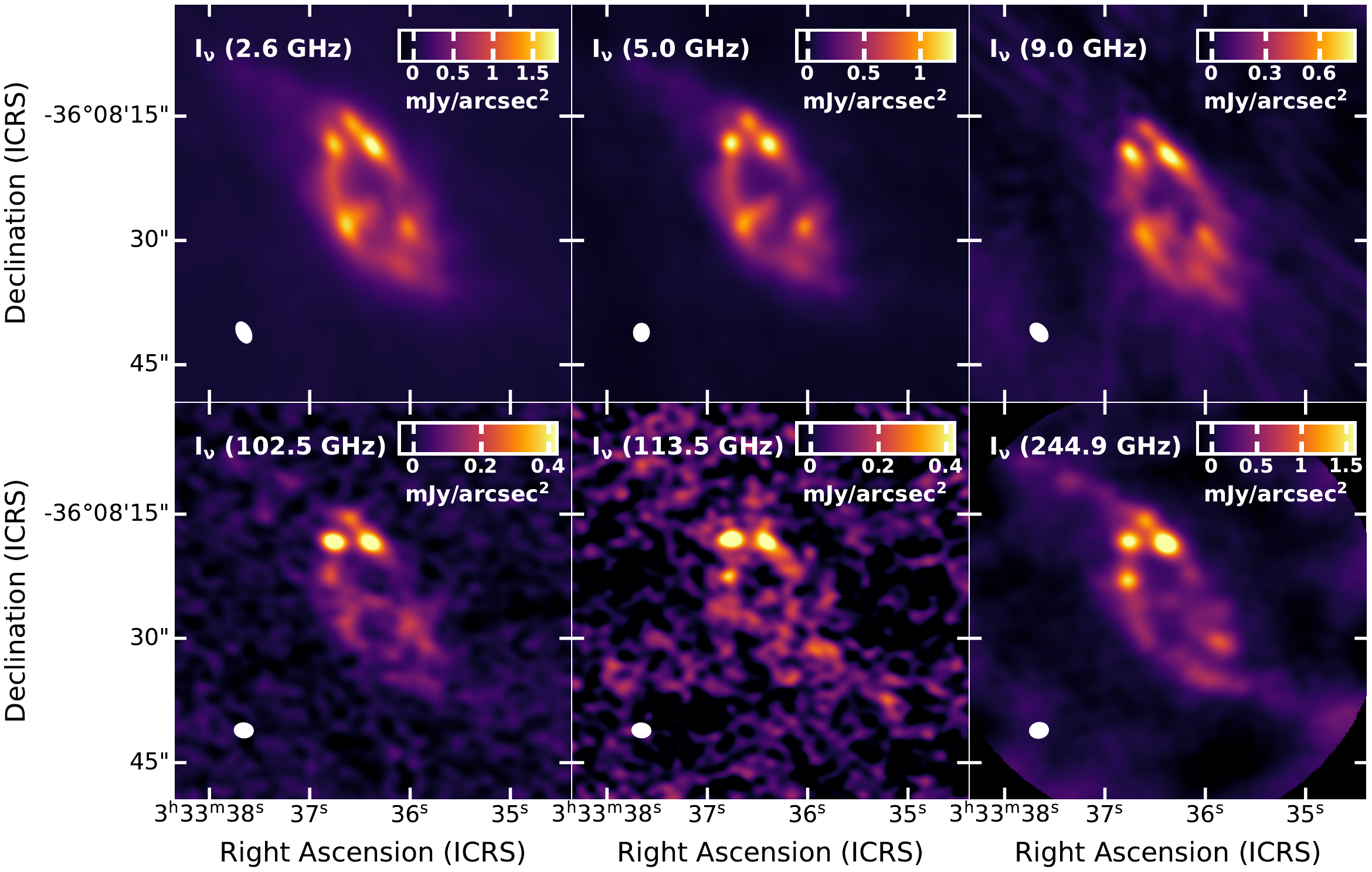} 
  \caption{Images of the VLA 2.6, 5.0, 9.0~GHz and ALMA 102.5, 113.5 and 244.9~GHz continuum in the central 48 arcsec of NGC~1365.  In each panel, the restoring beam is illustrated as the filled white ellipse at the lower-left side.}
  \label{fig:Image}
\end{figure*}

The 2.6 to 244.9~GHz images of the central 48~arcsec of NGC~1365 are displayed in Figure~\ref{fig:Image}. The 2.6~GHz, 5~GHz and 9~GHz continuum are mainly dominated by synchrotron radiation (see more in Section \ref{subsec:SED_analysis}) that originates from the AGN and supernova remnants.  These three images show that the presence of synchrotron emission from supernovae in the circumnuclear ring, particularly in the N2, N3 and E2 regions, although fainter emission is seen from the N1 and S regions.  Low frequency continuum emission is also detected from the AGN in the C region, but it is fainter than the sources in the circumnuclear ring. The C and E2 regions exhibit an outflow or jet-like feature, which is consistent with previous radio and X-ray observations \citep[e.g.,][]{Sandqvist1995,Wang2009,Venturi2017}.
The 102.5~GHz and 113.5~GHz continuum emission is mainly produced by photoionized gas associated with star formation (Section \ref{subsec:SED_analysis}).  These two images show that the star formation in the circumnuclear ring is concentrated in the N2 and N3 regions, which also harbor several massive young star clusters \citep[e.g.,][]{Whitmore2023}.  We also see fainter free-free emission from the N1, E1 and S regions.
The 244.9~GHz continuum emission is dominated by thermal dust emission (Section \ref{subsec:SED_analysis}).  At this frequency, the brightnesses of the various regions within the circumnuclear ring are similar to what is seen in the 102.5~GHz and 113.5~GHz images.  This is probably because the dust is very closely associated with the star-forming regions.  However, the E1 region is notably bright relative to other sources at 244.9~GHz, indicating that it is more dusty.

\section{Spectral Energy Distribution} \label{sec:SED}

\subsection{SED fitting} \label{subsec:SED_fit}

\begin{table*}
\centering
\begin{minipage}{0.9\textwidth}
\caption{Continuum measurements for SED analysis.}\label{tab:SED_flux}
    \begin{tabular}{@{}cccccccccccc@{}}
    \hline
    \hline
      Rest frame &
      \multicolumn{10}{c}{Flux density} 
      \\
      frequency &
      T region & 
      N region & 
      N1 region & 
      N2 region & 
      N3 region & 
      E region &  
      E1 region &  
      E2 region & 
      C region & 
      S region \\
      (GHz) &
      (mJy) &
      (mJy) &
      (mJy) &
      (mJy) &
      (mJy) &
      (mJy) &
      (mJy) &
      (mJy) &
      (mJy) &
      (mJy) \\
    \hline
2.6 & 223 $\pm$ 22 & 64 $\pm$ 6 & 11.9 $\pm$ 1.2 & 13.1 $\pm$ 1.3 & 15.7 $\pm$ 1.6 & 39 $\pm$ 4 & 9.6 $\pm$ 1.0 & 22 $\pm$ 2 & 7.9 $\pm$ 0.8 & 32 $\pm$ 3 \\
3.6 & 173 $\pm$ 17 & 52 $\pm$ 5 & 9.5 $\pm$ 0.9 & 11.2 $\pm$ 1.1 & 13.1 $\pm$ 1.3 & 30 $\pm$ 3 & 7.3 $\pm$ 0.7 & 17.0 $\pm$ 1.7 & 6.3 $\pm$ 0.6 & 25 $\pm$ 2 \\
5.0 & 131 $\pm$ 13 & 42 $\pm$ 4 & 8.0 $\pm$ 0.8 & 9.6 $\pm$ 1.0 & 10.6 $\pm$ 1.1 & 25 $\pm$ 3 & 6.1 $\pm$ 0.6 & 13.6 $\pm$ 1.4 & 5.2 $\pm$ 0.5 & 20 $\pm$ 2 \\
7.0 & 99 $\pm$ 10 & 32 $\pm$ 3 & 6.3 $\pm$ 0.6 & 7.8 $\pm$ 0.8 & 8.4 $\pm$ 0.8 & 19 $\pm$ 2 & 4.7 $\pm$ 0.5 & 10.6 $\pm$ 1.1 & 4.0 $\pm$ 0.4 & 15.4 $\pm$ 1.5 \\
9.0 & 82 $\pm$ 8 & 24 $\pm$ 2 & 4.4 $\pm$ 0.4 & 6.1 $\pm$ 0.6 & 6.6 $\pm$ 0.7 & 14.4 $\pm$ 1.4 & 3.3 $\pm$ 0.3 & 8.6 $\pm$ 0.9 & 2.9 $\pm$ 0.3 & 14.2 $\pm$ 1.4 \\
103.1 & 28.0 $\pm$ 1.4 & 11.9 $\pm$ 0.6 & 2.1 $\pm$ 0.1 & 3.5 $\pm$ 0.2 & 3.7 $\pm$ 0.2 & 5.1 $\pm$ 0.3 & 1.7 $\pm$ 0.1 & 2.4 $\pm$ 0.1 & 1.2 $\pm$ 0.1 & 5.0 $\pm$ 0.3 \\
114.1 & 31.2 $\pm$ 1.6 & 13.1 $\pm$ 0.7 & 1.6 $\pm$ 0.1 & 4.0 $\pm$ 0.2 & 3.6 $\pm$ 0.2 & 4.9 $\pm$ 0.3 & 2.0 $\pm$ 0.1 & 1.9 $\pm$ 0.1 & 1.0 $\pm$ 0.1 & 4.9 $\pm$ 0.3 \\
231.8 & 107 $\pm$ 11 & 44 $\pm$ 4 & 8.0 $\pm$ 0.8 & 10.3 $\pm$ 1.0 & 13.5 $\pm$ 1.4 & 20 $\pm$ 2 & 8.7 $\pm$ 0.9 & 7.8 $\pm$ 0.8 & 3.5 $\pm$ 0.4 & 16.7 $\pm$ 1.7 \\
246.3 & 129 $\pm$ 13 & 52 $\pm$ 5 & 9.7 $\pm$ 1.0 & 12.2 $\pm$ 1.2 & 16.1 $\pm$ 1.6 & 24 $\pm$ 2 & 10.3 $\pm$ 1.0 & 9.7 $\pm$ 1.0 & 4.0 $\pm$ 0.4 & 20 $\pm$ 2 \\

    \hline
    \end{tabular}
\end{minipage}
\end{table*}

For our analysis, we use all the ALMA and VLA continuum data between 2 and 246 GHz in the regions shown in Figure \ref{fig:Image_Aper} to construct SEDs. The continuum measurements used in our SED analysis are listed in Table \ref{tab:SED_flux}.  We fit the SEDs with functions representing synchrotron emission, free-free emission, and thermal dust emission.  The equation describing the overall emission is given as
\begin{equation}
\begin{split}
  f_\nu = {}&f_{\nu}(\mathrm{syn}) + f_\nu(\mathrm{ff}) + f_{\nu}(\mathrm{dust}) \\
={}&A_{\mathrm{syn}} \left(\frac{\nu}{1\ \mathrm{GHz}}\right)^{\alpha_\mathrm{syn}}
+A_{\mathrm{ff}} g_{\mathrm{ff}} + A_{\mathrm{dust}} \left(\frac{\nu}{200\ \mathrm{GHz}}\right)^{\alpha_\mathrm{dust}}
\end{split}
\label{eq:SED}
\end{equation}
where the synchrotron emission and thermal dust emission are treated as power laws and where the scales ($A_{\mathrm{syn}}$ and $A_{\mathrm{dust}}$) of these components as well as the scale of the free-free emission ($A_{\mathrm{ff}}$) are also free parameters (only set to be non-negative, adhering to physical priors).  The spectral index of the synchrotron emission ($\alpha_\mathrm{syn}$) is additionally treated as a free parameter.  Following a separate analysis to derived the spectra slope of the dust emission using ALMA Band 6, 7 and 8 data (which is described in Appendix \ref{appendix:dust}), we fixed the spectral index of the thermal dust emission ($\alpha_\mathrm{dust}$) to 4.0, which is approximately equivalent to $\beta=2.0$.
We use the Gaunt factor ($g_{\mathrm{ff}}$) derived for $hv \ll kT_e$ given by \citet{Draine2011} to set the shape of the free–free emission.  This function is given as
\begin{equation}
  g_{\mathrm{ff}} = 0.5535 \ln \left| 
  \left[ \frac{T_e}{\mathrm{K}} \right]^{1.5}
  \left[ \frac{\nu}{\mathrm{GHz}} \right]^{-1}
  Z^{-1}
  \right| -1.682,  
\label{eq:gff}
\end{equation} 
where $\nu$ is the frequency and $Z$ is the charge of the ions, which is set to 1.  For fitting the SED, we set $T_e$ to 7500~K, which is derived by the average value from the $\hii$ regions using auroral lines given by \cite{Bresolin2005}.   This value is slightly larger than those in other nearby starburst galaxies \citep[e.g.,][]{Bendo2015, Chen2023}; our measurements of the flux density from free-free emission varies by $\leq 10$ per cent when $T_e$ varies from 3000 to 10000~K.  

\setlength{\tabcolsep}{4.5pt}

\begin{table}
\centering
\begin{minipage}{0.5\textwidth}
\caption{SED analysis results.}\label{tab:SED_result}
    \begin{tabular}{@{}cccccc@{}}
    \hline
    \hline
       &
      \multicolumn{2}{c}{Synchrotron emission} &
      \multicolumn{1}{c}{Free-free emission} &
      \multicolumn{1}{c}{Dust emission} 
      \\
      Region &
      Spectra & 
      Fraction at & 
      Fraction at &
      Fraction at \\
       &
      index &
      103~GHz &
      103~GHz & 
      103~GHz \\
       &
       &
      (per cent) &
      (per cent) & 
      (per cent) \\
    \hline
      T & $-1.06 \pm 0.17$ & $13 \pm 8$ 
          & $75 \pm 8$ 
        & $12 \pm 1$ \\
      N & $-1.19 \pm 0.21$ & $5 \pm 4$ 
         & $84 \pm 4$ 
        & $11 \pm 1$ \\
      N1 & $-0.92 \pm 0.18$ & $19 \pm 13$ 
         & $67 \pm 13$ 
        & $14 \pm 1$ \\
      N2 & $-1.21 \pm 0.29$ & $3 \pm 3$ 
         & $89 \pm 4$ 
        & $8 \pm 1$ \\
      N3 & $-1.15\pm 0.23$ & $5 \pm 4$ 
         & $83 \pm 5$ 
        & $12 \pm 1$ \\
      E & $-0.94 \pm 0.17$ & $22 \pm 13$ 
         & $65 \pm 13$ 
        & $13 \pm 1$ \\
      E1 & $-1.23 \pm 0.19$ & $5 \pm 3$ 
         & $79 \pm 4$ 
        & $16 \pm 2$ \\
      E2 & $-0.80 \pm 0.12$ & $50 \pm 21$ 
         & $38 \pm 21$ 
        & $12 \pm 1$ \\
      C & $-0.98 \pm 0.18$ & $17 \pm 12$ 
         & $74 \pm 12$ 
        & $9 \pm 1$ \\
      S & $-0.87 \pm 0.18$ & $22 \pm 15$ 
         & $67 \pm 15$ 
        & $11 \pm 1$ \\
    \hline
    \end{tabular}
\end{minipage}
\end{table}

\setlength{\tabcolsep}{6pt}


With the {\sc lmfit} package (\citealt{Newville2022}), we first fit the SED in logarithmic space using the Levenberg-Marquardt algorithm, and then we sampled the posterior distributions of the fitting parameters by using the maximum likelihood via Markov chain Monte Carlo method in 1000 iterations.  From these parameters, we calculated the fractions of synchrotron emission, free-free emission, and thermal dust emission relative to the total emission.  Our measurements, along with their associated uncertainties, were derived from the medians and half the difference between the inner 68 per cent of the posterior distributions.


\subsection{SED analysis} \label{subsec:SED_analysis}


\begin{figure}
  \centering
  \includegraphics[width=0.45\textwidth]{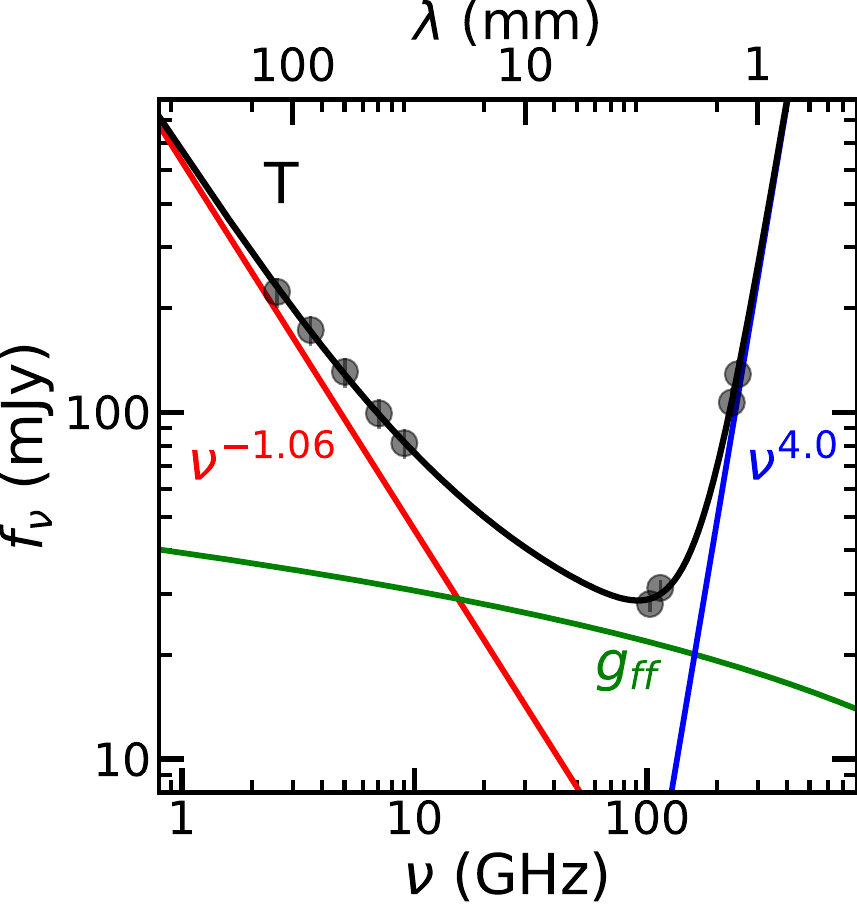} 
  \caption{The SED for the T region of NGC~1365 illustrated in Figure \ref{fig:Image_Aper}.  The black lines show the best fitting SED model for the data.  The red line represents synchrotron emission.  The green line represents free–free emission based on the Gaunt factor with $T_e$ fixed at 7500~K.  The blue line corresponds to thermal dust emission.}
  \label{fig:SED_T}
\end{figure}

\begin{figure*}
  \centering
  \includegraphics[width=0.85\textwidth]{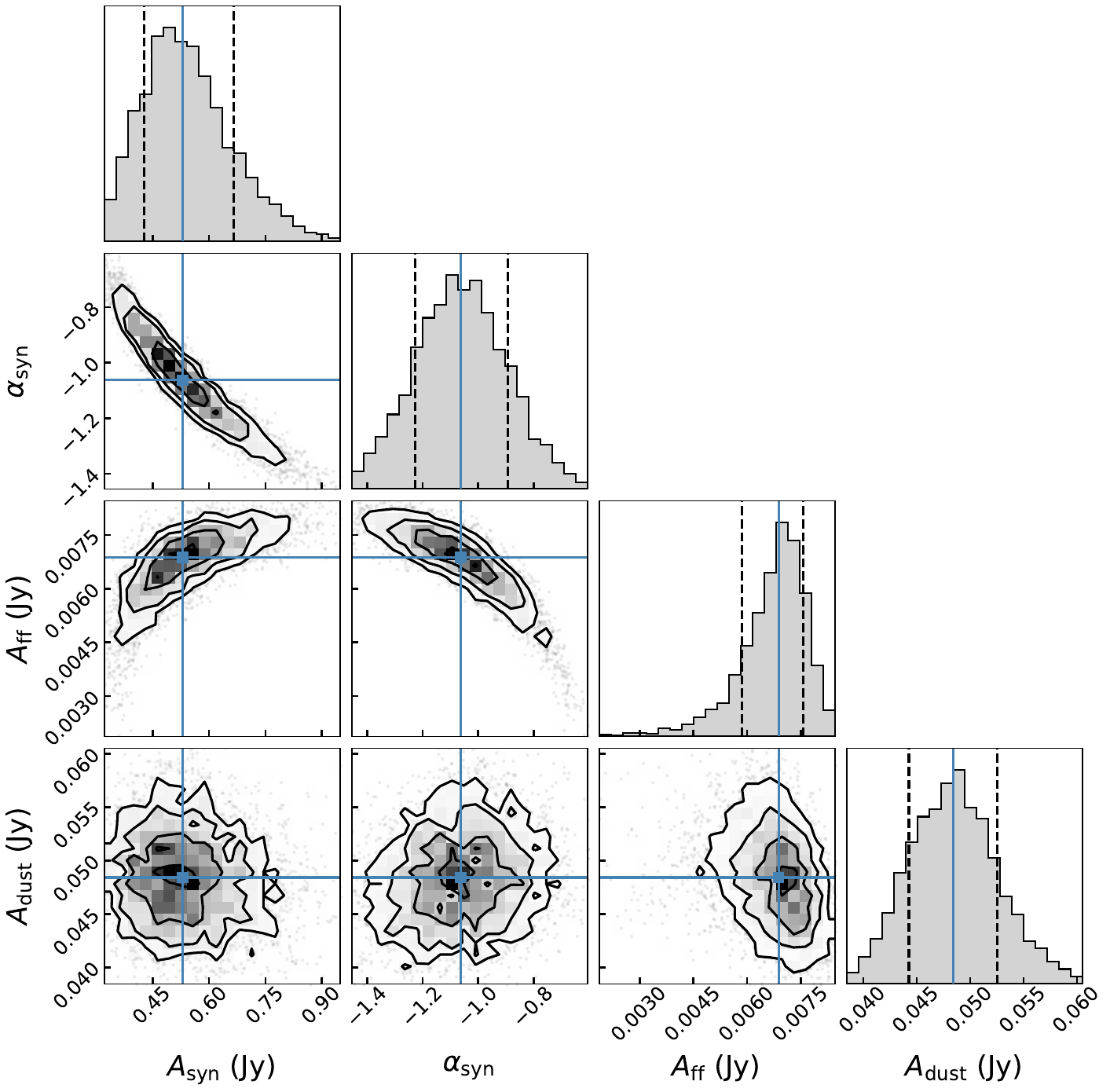} 
  \caption{The corner plot showcasing the posterior distributions for parameters characterizing the SED in Figure \ref{fig:SED_T}. In each 2-dimensional panel, solid contours enclose central 1, 2, and 3$\sigma$ levels of data points, with the median value denoted by a blue square intersected by two solid lines. In each 1-dimensional histogram, vertical dashed lines represent the 16th and 84th percentile values.}
  \label{fig:SED_T_corner}
\end{figure*}

\begin{figure*}
  \centering
  \includegraphics[width=1.0\textwidth]{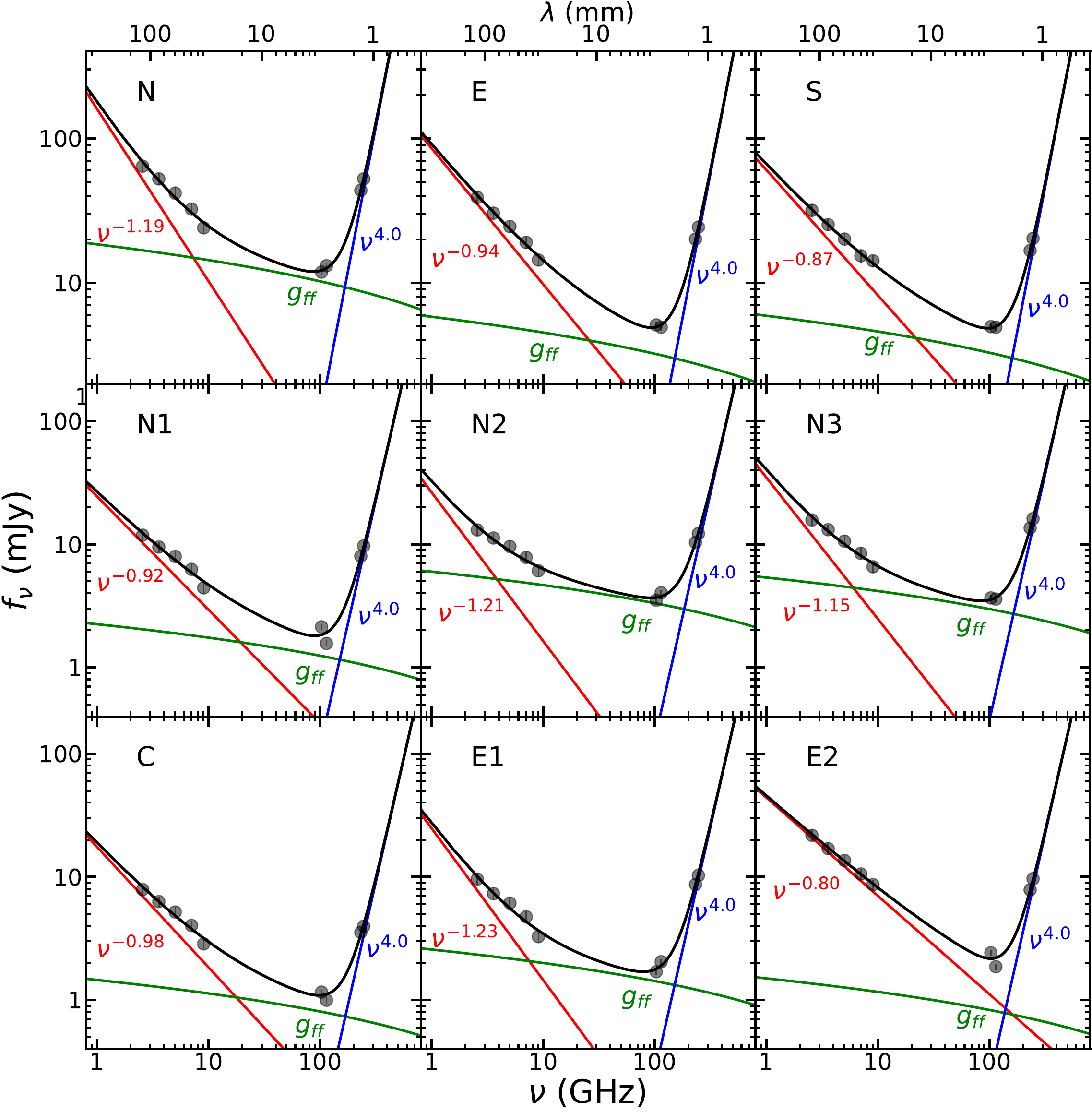} 
  \caption{The SEDs for the subregions of the T region illustrated in Figure \ref{fig:Image_Aper}. The lines with different colors are the same as those used in Figure \ref{fig:SED_T}.}
  \label{fig:SED}
\end{figure*}

The best fitting parameters from the SED fits are listed in Table \ref{tab:SED_result}, and the SEDs, along with the best-fitting functions, are shown in Figure \ref{fig:SED_T} and \ref{fig:SED}.
In the central 2.6 kpc diameter region (T region), the best fitting SED model reveals that free-free emission dominates between 30 and 150 GHz and produces 75 per cent of the total continuum emission at 103~GHz, while the remaining 25 per cent of 103~GHz continuum emission is produced by the synchrotron and dust emission.
The posterior distributions and correlation diagrams for parameters characterizing the SED are shown in Figure \ref{fig:SED_T_corner}. The scale of the free-free emission ($A_{\mathrm{ff}}$) is correlated with the spectral index of synchrotron emission ($\alpha_\mathrm{syn}$), but the variation of $A_{\mathrm{ff}}$ with $\alpha_\mathrm{syn}$ is not significant; $A_{\mathrm{ff}}$ ranges from 0.005 to 0.008 with $\alpha_\mathrm{syn}$ varying from $-1.4$ to $-0.8$.
The free-free emission also dominates the millimetre continuum emission in the N and S regions, while the fractions of the 103~GHz continuum emission originating from free-free emission vary between $\sim$ 60 and $\sim$ 90 per cent between these regions and among their subregions.  

The SED model for the central 380~pc diameter region (C region) shows that the free-free emission is dominated at the 103~GHz continuum emission as well, which seems unexpected given that the AGN within this region would be expected to produce significantly stronger synchrotron emission rather than free-free emission.  
Notably, significant broad Pa$\alpha$ line emission, which traces the same photoionized gas as free-free emission, is detected within the nucleus \citep{Fazeli2019}, though the emission at near-infrared wavelengths is still affected by dust obscuration.  Nonetheless, we can use the measured Pa$\alpha$ line emission to calculate the expected free-free emission in millimetre bands, which we can then compare to our SED. 
To calculate the Gaunt factor, we used 
\begin{equation}
  g_{\mathrm{ff}} = \ln \left|
  {0.1907
  \left\lgroup
  \left[ \frac{T_e}{\mathrm{K}}\right]^{1.5} 
  \left[ \frac{\nu}{\mathrm{GHz}}\right]^{-1} 
  \mathrm{Z}^{-1}
  \right\rgroup^{0.5513}}
  + 2.718 \right|,
\label{eq:gff_Pa}
\end{equation}
given by \citet{Draine2011} for situations where the frequency of Pa$\alpha$ line is comparable to $kT_e/h$.  We then derived the free–free emission using 
\begin{equation}
\begin{split}
\left[\frac{f_\nu(\mathrm{ff})}{\mathrm{Jy}} \right]=
\left[\frac{\int{f_\nu(\mathrm{Pa}\alpha) dv}}{\mathrm{Jy km s}^{-1}}\right] 
  \exp\left\lgroup -0.048\left[\frac{\nu}{\mathrm{GHz}}\right]\left[\frac{T_e}{\mathrm{K}}\right]^{-1}\right\rgroup   
 \\
\times \left\lgroup4.38\times10^{33} g_{\mathrm{ff}}^{-1}  
  \left[\frac{\epsilon_\nu}{\mathrm{erg s}^{-1}\mathrm{ cm}^{-3}}\right]
  \left[\frac{\nu}{\mathrm{~GHz}}\right]^{-1}
  \left[\frac{T_e}{\mathrm{~K}}\right]^{0.5} 
\right\rgroup^{-1},
\label{eq:ff_Pa}
\end{split}
\end{equation}
based on equations from \citet{Draine2011} and \citet{Scoville2013} where $\epsilon_\nu$ is the emissivity from \citet{Storey1995}, the hydrogen recombination line emission integrated over velocity is represented by $\int{f_\nu(\mathrm{Pa}\alpha) dv}$ and the free–free emission is represented by $f_\nu$(ff).
Using the extinction-corrected Pa$\alpha$ flux of $2.45\times10^{-13}$ erg cm$^{-2}$ s$^{-1}$ as calculated from the measurements from \citet{Fazeli2019} and assuming an electron temperature of 7500~K, we derived the free-free flux density of 1.4 mJy at 103~GHz. Although this measurement is 1.5 times larger than our measurement of 0.9 mJy, it is crucial to note that the free-free flux density derived from the Pa$\alpha$ line is dependent on $T_e$.  If we use an electron temperature of 6000~K, the free-free flux density at 103~GHz derived from the Pa$\alpha$ line would be is 0.9 mJy, which would match our measurement.  
This comparison also indicates that exotic forms of emission, such as ``submillmetre excess emission'', do not significantly affect the overall flux densities measured in ALMA Band 3 (This is discussed further in Appendix~\ref{appendix:dust}).
To summarize, this calculation demonstrates that the level of free-free emission that we are measuring from the nuclear region is indeed consistent with the previously observed Pa$\alpha$ line emission.

As an additional note about the C region, the thermal dust emission within this region, indicated by the 240~GHz flux density, is relatively low compared to other regions within the central starburst ring.    
\citet{Combes2019} demonstrated that the 350~GHz continuum emission within the C region is concentrated within a torus with a diameter of $\sim 50$ pc.  The dust is relatively hot ($\sim$1300 K; \citealt{Fazeli2019}), but apparently the dust mass is very low, thus leading to the relatively low thermal dust emission in the submillimetre and millimetre bands.


In the E region, the free-free emission dominates between 50 and 120~GHz, but only produces 65 per cent of the 103~GHz continuum emission, while the synchrotron and dust emission produce the remaining 35 per cent continuum emission.  This region can be divided into two distinct subregions with very different SEDs.  Within the E1 region, $\sim 80$ per cent of the 103~GHz emission originates from free-free emission, which likely arises mainly from the two newly identified massive young star clusters in this region \citep[][]{Whitmore2023}.  In the neighboring E2 region, however, the synchrotron emission produces approximately half of the 103~GHz continuum emission.  This enhanced synchrotron emission may be attributed to outflows driven by the AGN or starburst as also observed in $\oiii$ emission \citep[e.g.,][]{Venturi2017}, X-ray emission \citep[e.g.,][]{Wang2009} and radio emission (see, for example, \citealt{Sandqvist1995} as well as the top panels of Figure \ref{fig:Image}), although morphology of these outflows varies in different wavebands.  Despite the presence of young star clusters within the E2 region as noted by \citet[][]{Whitmore2023}, the free-free emission from these clusters is still relatively weak in comparison to the synchrotron emission. 

Above all, the radio-to-submillimetre SEDs of the pseudobulge of NGC1365 can be well characterised by the combination of synchrotron emission, free-free emission, and thermal dust emission.  However, these SED fits have not included either submillimetre excess emission, which has been detected in the 230 to 350~GHz frequency range in some nearby galaxies \citep[e.g.,][]{Galliano2003,Galliano2005,Bendo2006,Galametz2009,Galliano2011,Planck2011_XVII,Hermelo2016,Chang2020}, or anomalous microwave emission (AME), which has been detected in the 20 to 60~GHz range in other nearby galaxies \citep[e.g.,][]{Murphy2010,Hensley2015,Planck2015,Murphy2018,Battistelli2019,Harper2023,Fernandez2024}.  To investigate the possible contributions of these forms of emission to the SED, we performed additional analyses on the SED of NGC~1365 in Appendices \ref{appendix:dust} and \ref{appendix:ame}.  We ultimately determined that these sources of emission are generally negligible compared to the synchrotron, free-free, and thermal dust components used in our SED fits.




\subsection{Comparisons of SEDs to previous studies}

The spectral indices of the radio continuum emission in the central pseudobulge of NGC~1365 were previously measured by \citet{Sandqvist1995} using VLA 20 cm (1.5 GHz), 6 cm (4.9 GHz) and 2 cm (14.9 GHz) data with a resolution of 2.36 $\times$ 0.95 arcsec$^2$ and by \citet{Stevens1999} using the Australia Telescope Compact Array 3 cm and 6 cm data with a 2 arcsec diameter aperture. Despite variations in our data and methodology, our measured spectral index of $-0.98\pm0.18$ in the C region from our SED analysis falls within the range of their measurements, which were $-0.87$ and $-1.1$ in the nucleus. Similarly, the spectral index of $-0.80\pm0.12$ in the E2 region from our SED analysis aligns with their measurements of $-0.94$ and $-1.0$ in the comparable regions.   
However, their spectral indices for the N2 and N3 regions, which range from $-0.6$ to $0.07$,  are notably larger than the synchrotron spectral indices of $-1.21$ and $-1.15$ from our SED analysis.  This disparity arises simply because the spectral indices measured by \citet{Sandqvist1995} and \citet{Stevens1999} are empirical measurements of the spectral slope that are not based on any spectral decomposition and therefore reflect the blending of synchrotron and free-free emission in the SED at these frequencies.   
Without any spectral decomposition, the spectral indices based on our data from 2.6 to 9~GHz for the N2 and N3 regions are $-0.59 \pm 0.05$ and $-0.68 \pm -0.04$, respectively. These values fall at the lower limit of the range of \citet{Sandqvist1995} and \citet{Stevens1999} measurements.

The slopes of synchrotron radiation exhibit a range between $-1.59$ and $-0.74$ among different nearby galaxies \citep[][]{Peel2011,Bendo2015,Bendo2016,Salak2017,Michiyama2020,Chen2023}.  However, despite these variations, the free-free emission has been found to still dominate the continuum emission within the frequency range of 50--150 GHz in other galaxies \citep[][]{Peel2011,Bendo2015,Bendo2016,Salak2017,Michiyama2020,Chen2023}. 
Our SED result of the central pseudobulge (T region) in NGC~1365 is consistent with these findings, as shown in Figure \ref{fig:SED_T}.  Nonetheless, the fractions of millimetre continuum emission originating from free-free emission varies significantly across different regions, as illustrated in Figure \ref{fig:SED}.  The fraction of the 103~GHz emission originating from free-free emission varies from 38 to 89 per cent among the subregions that we examined within the pseudobulge, and the dominant source of millimetre continuum emission may shift from free-free emission to synchrotron emission.
Smaller variations in the fraction of millimetre emission originating from free-free emission have also been observed between the two neighboring nuclei of NGC~3256 \citep[from 76 to 90 per cent;][]{Michiyama2020}, and between the nucleus and the second brightest star-forming region of NGC~1808 \citep[from 62 to 86 per cent;][]{Chen2023}. In both cases, however, the fraction of millimeter continuum emission originating from free-free emission is still significantly higher than that from synchrotron radiation, which differs from what we see in the E2 region.
Synchrotron emission that is notably strong relative to the free-free emission from star-forming regions would only be expected in situations where an AGN jet is the dominant energy source and where star formation is relatively weak or largely absent.  

Notably, \citet{Michiyama2023} published an analysis of the SED of the AGN in NGC 1068 in which they found emission associated with an X-ray corona that did not follow a standard synchrotron power law but exhibited significant self-absorption at $\gtsim 20$~GHz.  However, we do not observe any such phenomena in any of the regions where we extracted SEDs, including the AGN, although we may need additional data between 10 and 100~GHz to properly detect this type of coronal emission.




\section{Star Formation Rates} \label{sec:SFR}

\subsection{SFR measurements from free-free emission} \label{subsec:SFR_ALMA}

The SFR can be calculated from the photoionizing photon production rate ($Q$) using
\begin{equation}
  \frac{\mathrm{SFR}}{\mathrm{M}_\odot~\mathrm{yr}^{-1}}
   =7.29\times10^{-54}\frac{Q}{\mathrm{s}^{-1}},
\label{eq:QtoSFR}
\end{equation}
where the conversion was derived from \citet{Murphy2011} using the {\sc Starburst99} model \citep{Leitherer1999} with solar metallicity ($Z_\odot$ = 0.020) and a \citet{Kroupa2002} initial mass function for a mass range of 0.1--100~M$_\odot$.  While different conversions that account for the stellar rotation from \citet{Leitherer2014} were used in previous studies \citep{Bendo2015,Bendo2016,Bendo2017}, these conversions are not widely used to compare with the SFRs from other conversion equations, and thus we use a standard conversion that includes no stellar rotation effects to simplify our comparison of SFRs.

The free–free emission can be related to the values to $Q$ by applying
\begin{equation}
  Q=\alpha_B n_e n_p V
\label{e_emtoq}
\end{equation}
from \citet{Scoville2013}, where $n_e$ and $n_p$ represent the electron and proton densities.  $\alpha_B$ is the effective recombination coefficient from \citet{Storey1995}, which varies significantly with $T_e$, but remains relatively stable for $n_e$ values spanning from $10^2$ to $10^5$~cm$^{-3}$.  We therefore keep the electron density $n_e$=$10^3$~cm$^{-3}$ constant and use an $\alpha_B$ corresponding to the $T_e$ of 7500~K set in Section \ref{sec:SED}.   Following this, we can convert the free-free flux density, $f_\nu$(ff), to SFRs using
\begin{equation}
\begin{split}
\frac{\mathrm{SFR}(\mathrm{ff})}{\mathrm{M}_\odot\mathrm{ yr}^{-1}}=
  1.28\times10^{11}
  \ \ \ \ \ \ \ \ \ \ \ \ \ \ \ \ \ \ \ \ \ \ \ \ \ \ \ \ \ \ \ \ \ \ \ \ \ \ \ \ \ \ \ \ \ \ \ \ \ \ \ \ \\
  \times g_{\mathrm{ff}}^{-1}
  \left[\frac{\alpha_B}{\mathrm{ cm}^3\mathrm{ s}^{-1}}\right]
  \left[\frac{T_e}{\mathrm{~K}}\right]^{0.5}
  \left[\frac{D}{\mathrm{ Mpc}}\right]^{2}
  \left[\frac{f_\nu(\mathrm{ff})}{\mathrm{Jy}}\right],
\end{split}
\label{eq:SFR_freefree}
\end{equation}
where $D$ is the distance of 19.6~Mpc.  Notably, the SFRs are dependent upon the values of $T_e$, which we have set to 7500~K; the SFRs may vary by $\sim30$ per cent from our values for $T_e$ ranging from 5000~K to 15000~K.

\begin{table}
\centering
\begin{minipage}{0.435\textwidth}
\caption{SFR measurements from 103 GHz free-free emission.}\label{tab:SFR}
    \begin{tabular}{@{}ccccccc@{}}
    \hline
    \hline
    Region &
    $f_\nu(\mbox{ff})$&
    SFR &
    Area &
    $\Sigma_{\rm SFR}$ \\ 
    & 
    (mJy)& 
    (M$_\odot$ yr$^{-1}$) & 
    (kpc$^2$) & 
    (M$_\odot$ yr$^{-1}$kpc$^{-2}$) \\
    \hline
T & 20.2 $\pm$ 1.1 & 8.9 $\pm$ 1.1 & 5.56 & 1.6 $\pm$ 0.2 \\
N & 9.9 $\pm$ 0.5 & 4.4 $\pm$ 0.3 & 0.74 & 5.9 $\pm$ 0.4 \\
N1 & 1.44 $\pm$ 0.07 & 0.63 $\pm$ 0.13 & 0.11 & 5.6 $\pm$ 1.2 \\
N2 & 3.14 $\pm$ 0.16 & 1.39 $\pm$ 0.09 & 0.11 & 12.2 $\pm$ 0.8 \\
N3 & 3.06 $\pm$ 0.15 & 1.35 $\pm$ 0.10 & 0.11 & 11.9 $\pm$ 0.9 \\
E & 3.33 $\pm$ 0.17 & 1.5 $\pm$ 0.3 & 0.43 & 3.5 $\pm$ 0.7 \\
E1 & 1.34 $\pm$ 0.07 & 0.59 $\pm$ 0.04 & 0.11 & 5.2 $\pm$ 0.4 \\
E2 & 0.92 $\pm$ 0.05 & 0.4 $\pm$ 0.2 & 0.20 & 2.0 $\pm$ 1.2 \\
C & 0.85 $\pm$ 0.04 & 0.38 $\pm$ 0.06 & 0.11 & 3.3 $\pm$ 0.6 \\
S & 3.34 $\pm$ 0.17 & 1.5 $\pm$ 0.4 & 0.45 & 3.2 $\pm$ 0.8 \\
    \hline
    \end{tabular}
\end{minipage}
\end{table}

The SFRs measured from the 103~GHz free-free emission (i.e., the 103~GHz continuum emission multiplied by the free-free fractions at 103~GHz listed in Table~\ref{tab:SED_result}) are listed in Table~\ref{tab:SFR}.  Since the free-free emission from the C region is mainly related to the AGN (as discussed in Section \ref{subsec:SED_analysis}), the SFRs for the T region are calculated after subtracting the emission from the C region.  These values are used for comparisons to the SFRs obtained from other bands.  

The resulting SFR from the central pseudobulge (excluding the AGN) is $8.9 \pm 1.1$~M$_\odot$ yr$^{-1}$. 
Remarkably, nearly half of the star formation is concentrated within the N region, which occupies only 13 per cent of the area of the T region.  The star formation in the N region is predominantly within two bright sources (N2, N3) where the star formation surface densities reach 12 M$_\odot$ yr$^{-1}$kpc$^{-2}$.  This is seven times larger than the star formation surface density averaged over the T region.  The SFRs of the E and S regions are both about 15 per cent of the entire pseudobulge, and the corresponding star formation surface densities are twice that of the T region.

\subsection{Comparisons of millimetre and previously-published H$\alpha$ SFRs}

\citet{Schinnerer2023} derived a SFR of 5~M$_\odot$ yr$^{-1}$ from the attenuation-corrected H$\alpha$ line emission within the central 3.6~kpc diameter region. Even though \citet{Schinnerer2023} used a larger aperture, their SFR is still lower than our measurement within our T region, probably because of inaccuracies in the dust attenuation correction applied to the $\ha$ data within this dusty starburst. 
The dust extinction within the central regions of NGC~1365 ranges from 4 to 9 \citep{Fazeli2019}, which indicates that the observed H$\alpha$ luminosity will only be 0.025 per cent to 2.5 per cent of the total H$\alpha$ emission from the stellar population.
A similar situation is also seen in other dusty starburst regions \citep[e.g.,][]{Chen2023}.


Many researchers have calculated SFRs using ultraviolet or $\ha$ emission added to rescaled versions of the infrared emission from dust measured in either a specific band or integrated over a large frequency range \citep[e.g.,][]{Kennicutt2012}.  The infrared emission in these situations represent an extinction correction to the ultraviolet or $\ha$ data.  However, given the large effect of dust extinction within dusty starbursts like NGC~1365, the ultraviolet and $\ha$ emission has been sufficiently obscured that the calculated SFRs end up being highly dependent on the infrared terms in these equations.  Hence, we will not attempt to calculate SFRs using combinations of ultraviolet or $\ha$ data with infrared data.


\subsection{Comparisons of millimetre and radio SFRs} \label{subsec:SFR_radio}

\begin{table}
\centering
\begin{minipage}{0.48\textwidth}
\caption{SFR measurements from 1.5 GHz emission.}\label{tab:SFR_radio}
    \begin{tabular}{@{}ccccc@{}}
    \hline
    \hline
    Region &
    $f_\nu(\mbox{1.5 GHz})$&
    SFR &
    SFR$-$SFR(ff) &
    $\Delta$SFR$^a$/SFR(ff)\\ 
    & 
    (mJy)& 
    (M$_\odot$ yr$^{-1}$) & 
    (M$_\odot$ yr$^{-1}$) & 
    (per cent) \\
    \hline
T & $285 \pm 30$ & $8.3 \pm 0.9$ & $-0.6 \pm 1.4$ & $-7 \pm 1$ \\
N & $85 \pm 9$ & $2.5 \pm 0.2$ & $-1.9 \pm 0.4$ & $-44 \pm 5$ \\
N1 & $15.1 \pm 1.5$ & $0.44 \pm 0.04$ & $-0.20 \pm 0.14$ & $-31 \pm 7$ \\
N2 & $17.0 \pm 1.7$ & $0.49 \pm 0.05$ & $-0.89 \pm 0.10$ & $-64 \pm 8$ \\
N3 & $19.5 \pm 1.9$ & $0.56 \pm 0.06$ & $-0.78 \pm 0.12$ & $-58 \pm 7$ \\
E & $55 \pm 6$ & $1.61 \pm 0.16$ & $0.14 \pm 0.35$ & $9 \pm 2$ \\
E1 & $14.2 \pm 1.4$ & $0.41 \pm 0.04$ & $-0.18 \pm 0.06$ & $-30 \pm 4$ \\
E2 & $30 \pm 3$ & $0.86 \pm 0.09$ & $0.5 \pm 0.2$ & $112 \pm 65$ \\
S & $46 \pm 5$ & $1.33 \pm 0.13$ & $-0.15 \pm 0.37$ & $-10 \pm 3$ \\
    \hline
\end{tabular}\\
$^a$The difference between the SFRs measured from 1.5 GHz radio emission and 103 GHz free-free emission.
\end{minipage}
\end{table}

For calculating SFRs based on the radio emission, we first measure the 1.5~GHz flux densities in the same apertures used in Section \ref{subsec:SFR_ALMA} from the VLA 1.5~GHz data described in Section \ref{subsec:VLA} (which also meant subtracting the emission from the C region from the emission within the T region). Following this, we convert the 1.5~GHz flux densities to SFRs using 
\begin{equation}
  \frac{\mathrm{SFR}(1.5\mathrm{GHz})}{\mathrm{M}_\odot\mathrm{ yr}^{-1}}
  = 0.0760 \left[\frac{D}{\mathrm{Mpc}}\right]^{2}
  \left[\frac{f_\nu(1.5 \mathrm{ GHz})}{\mathrm{Jy}}\right]
\label{eq:sfr_radio}
\end{equation}
given by \citet{Murphy2011}.  The SFRs measured from the 1.5 GHz emission are listed in Table~\ref{tab:SFR_radio}.  

\begin{figure}
  \centering
  \includegraphics[width=0.48\textwidth]{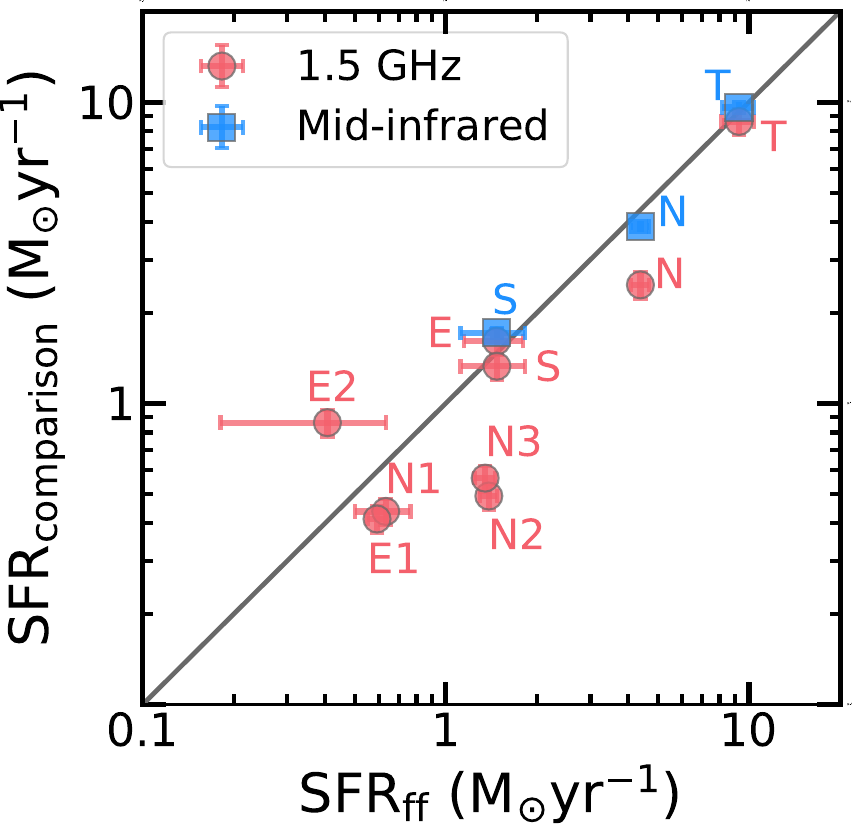} 
  \caption{The differences between the SFRs from free-free emission and from 1.5 GHz radio emission (red circles) and the difference between the SFRs from free-free emission and from 24~$\mu$m continuum emission (blue squares).}  
  \label{fig:sfr_delta}
\end{figure}

The difference between the SFRs measured from the 1.5 GHz radio emission and 103~GHz free-free emission is illustrated as filled circles in Figure \ref{fig:sfr_delta}.
The SFR averaged over the T region estimated from the 1.5~GHz emission is comparable to the SFR from the free-free emission; the relative discrepancy between them is 10 per cent. 
However, the 1.5~GHz emission in the N region yields a SFR much lower than the SFR measured from the free-free emission, particularly in its two starburst dominated subregions, N2 and N3.  This difference is likely attributed by the diffusion through the interstellar medium of the cosmic rays producing the synchrotron radiation \citep[e.g.,][]{Murphy2006}, which would make the radio emission appear more extended than other star formation tracers.  Since the T region is much larger than the N region, it covers both the star-forming regions producing the free-free emission and the more diffuse regions emitting the synchrotron emission associated with those regions, so the SFRs in the T region are more consistent than in the smaller N region.

The E1 region is very similar to the subregions of the N region in that the SFR from the 1.5~GHz emission is low relative to the 103~GHz free-free emission, which again is likely to be related to the diffusion of the electrons producing the synchrotron emission from the star-forming region.
Conversely, in the E2 region, the SFR measured from the 1.5~GHz emission is much higher than the SFRs derived from the free-free emission; the relative discrepancy between them is more than 100 per cent.  This is likely because the 1.5~GHz emission in the E2 region is not entirely related to star formation but includes significant emission from the outflow within that region (e.g., the top panels of Figure \ref{fig:Image}).  
The consistency in SFRs measured between the 1.5~GHz emission and free-free emission within the E region arises from mixing the emission from E1 and E2.

The difference in SFRs measured from the 1.5~GHz emission and free-free emission is not significant despite the fact that the S region is smaller than the N region and we would expect the electrons producing the synchrotron emission to diffuse outwards from the S region in the same way that they have seemed to diffuse outwards from the N region.  Given that the S region does not have an outflow like the E2 region, the consistency in SFRs derived from the 1.5~GHz emission and free-free emission within the S region may imply the presence of possibly more supernova remnants than in the N region.  However, we cannot draw this conclusion because of the lack of direct evidence of supernovae remnants within the S region.  Further high angular resolution radio observations of the pseudobulge would be instrumental in providing a clearer understanding of what is happening.


\subsection{Comparisons of millimetre and mid-infrared SFRs} \label{subsec:SFR_IR}

\begin{figure}
  \centering
  \includegraphics[width=0.48\textwidth]{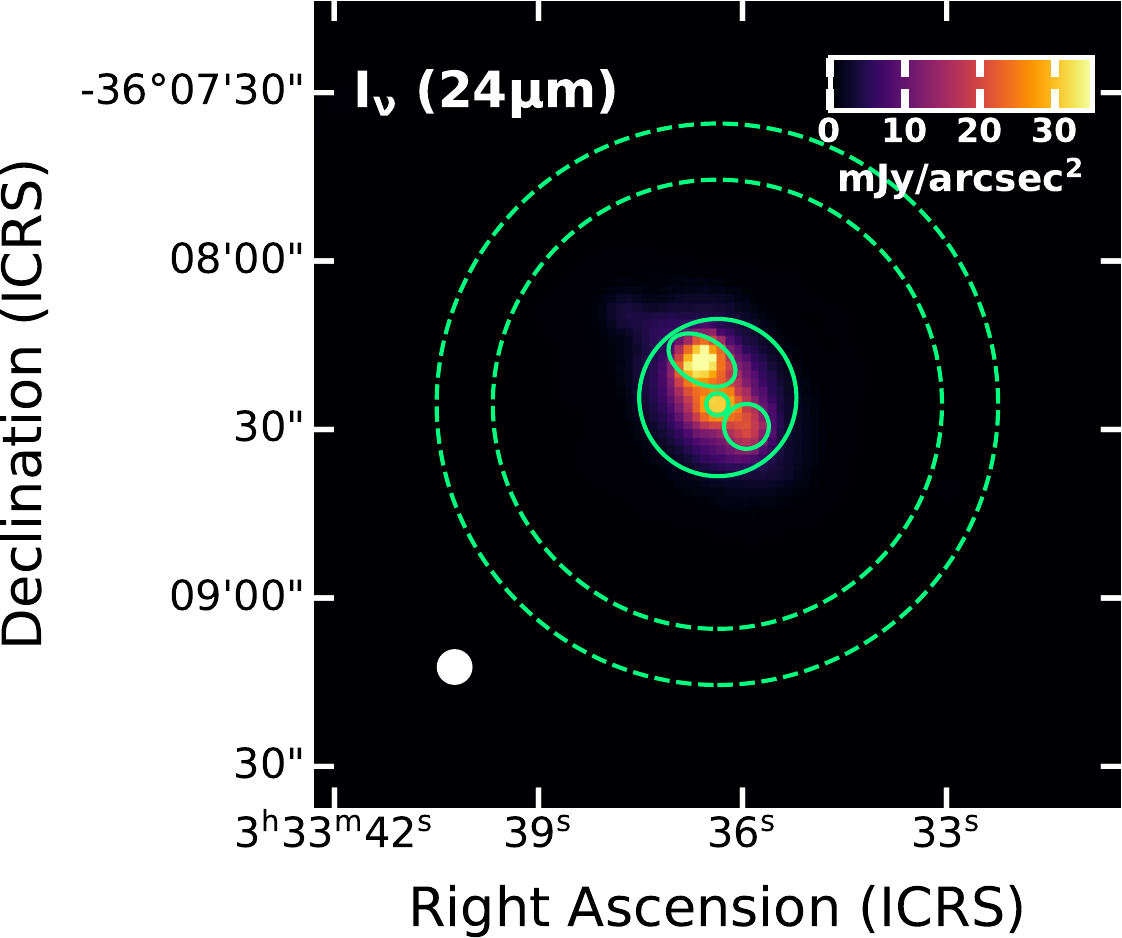} 
  \caption{The image of the {\it Spitzer} 24~$\mu$m emission in the central 144 arcsec of NGC~1365. The color is set as linear scale to show the infrared emission in the central pseudobulge region.  The solid green circles show the T, N, C and S regions, and the dashed green circles show the annuli used for the local background subtraction. The restoring beam is illustrated as the filled white ellipse at the lower-left side.}
  \label{fig:Image_IR}
\end{figure}

\setlength{\tabcolsep}{2.25pt}

\begin{table}
\centering
\begin{minipage}{0.48\textwidth}
\caption{SFR measurements from mid-infrared flux.}\label{tab:SFR_IR}
    \begin{tabular}{@{}cccccc@{}}
    \hline
    \hline
    Region &
    Aperture &
    $f_\nu(\mbox{MIR})$ &
    SFR &
    SFR$-$SFR(ff) &
    $\Delta$SFR$^a$$/$SFR(ff) \\  
    & 
    correction &
    (Jy)& 
    (M$_\odot$ yr$^{-1}$) & 
    (M$_\odot$ yr$^{-1}$) & 
    (per cent) \\
    \hline
T & 1.16 & $8.1 \pm 0.3$ & $9.6 \pm 0.4$ & $0.7 \pm 1.2$ & $8 \pm 1$ \\
N & 1.61 & $3.27 \pm 0.13$ & $3.86 \pm 0.15$ & $-0.5 \pm 0.4$ & $-12 \pm 1$ \\
S & 1.61 & $1.45 \pm 0.06$ & $1.71 \pm 0.07$ & $0.2 \pm 0.4$ & $16 \pm 4$ \\
\hline 
\end{tabular}\\
$^a$The difference between the SFRs measured from 24~$\mu$m emission and 103~GHz free-free emission.
\end{minipage}
\end{table}

\setlength{\tabcolsep}{6pt}

For this comparison, we used the {\it Spitzer} 24~$\mu$m image to derive the mid-infrared (MIR) SFRs for the central pseudobulge of NGC~1365. Given that the beam FWHM of {\it Spitzer} 24~$\mu$m image is larger than those of the ALMA images (as described in Section \ref{sec:data}), our comparison focuses on the SFRs from the T, N and S regions, which are larger than the beam FWHM of {\it Spitzer} 24~$\mu$m image. 

We first measured the flux densities in the T, N and S apertures, and then subtracted the flux density within the C region (which contains the AGN) from the T region.  We then subtracted the median background within an annulus with radii of 40--50 arcsec, and we applied aperture corrections given by \citet{Engelbracht2007}. 
The image of the {\it Spitzer} 24~$\mu$m emission associated with the apertures is shown in Figure \ref{fig:Image_IR}, and the aperture corrections factors and the corrected flux densities are listed in Table \ref{tab:SFR_IR}.
To convert {\it Spitzer} 24~$\mu$m flux densities to SFRs, we use
\begin{equation}
  \frac{\mathrm{SFR}(24\mu\mathrm{m})}{\mathrm{M}_\odot\mathrm{ yr}^{-1}}=
  2.44\times10^{-16} \left[\frac{D}{\mathrm{Mpc}}\right]^{2}
  \left[\frac{\nu f_\nu(24\mu\mathrm{m})}{\mathrm{Jy Hz}}\right]
\label{eq:sfr_24}
\end{equation}
given by \citet{Rieke2009}.  The SFRs measured in the T, N and S regions listed in Table \ref{tab:SFR_IR}.  
Notably, the corresponding MIR luminosities range from $2.2\times10^{9}$ to $1.2\times10^{10} \mathrm{L}_\odot$, falling within the applicable range of $6\times10^{8}$ to $1.3\times10^{10}\mathrm{L}_\odot$ for this relation.

The difference between the SFRs measured from 24~$\mu$m emission and 103~GHz free-free emission are shown as the filled squares in Figure \ref{fig:sfr_delta}. The relative discrepancy between the SFRs from the T, N and S regions estimated from the MIR emission is less than 1 $\sigma$. 
The conversions from MIR emission to SFRs are based on assumptions that the shape of infrared SED remains unchanged and that the monochromatic infrared emission scales linearly with the total infrared emission.
The strong correspondence between the 24~$\mu$m and 103~GHz SFRs indicate that these assumptions are applicable to the pseudobulge (excluding the AGN) within NGC~1365.  Note that the 24~$\mu$m flux density does not always yield the same SFR as measured from millimetre free-free emission.  In many low-metallicity dwarf galaxies, for example, the dust is unusually hot \citep{Hunt2005, Engelbracht2008, Rosenberg2008, Hirashita2009, Remy-Ruyer2013}, which affects the conversion of MIR flux densities to SFR and the comparison of SFRs from MIR flux densities to SFRs from other tracers \citep[e.g.,][]{Calzetti2010, Bendo2017}.  Conversely, \citet{Bendo2016} found that the dust densities in the AGN/starburst composite galaxy NGC 4945 were so high that the MIR emission was optically thick and therefore suppressed relative to millimetre star formation tracers.  Hence, it is still important to identify situations where SFRs from MIR emission correspond to SFRs from millimetre continuum emission, which is why it is important to show the agreement between these two SFRs in NGC 1365.


\subsection{Variations of free-free emission fractions with frequencies} \label{subsec:SFR_ff}

Free-free emission is a valuable SFR metrics because it can directly trace young stellar populations while remaining generally unaffected by dust attenuation \citep[e.g.,][]{Bendo2015,Bendo2016,Chen2023}.  Nevertheless, free-free emission is not widely used to measure SFRs.  This is not only because the millimetre continuum emission is so weak that only powerful millimetre instruments like ALMA can detect this form of emission from extragalactic sources but also because of the lack of comprehensive radio-to-submillimetre SED templates for identifying the fractions of continuum emission originating from the free-free emission.  Conducting SED analyses, as applied to this and other galaxies \citep[e.g.,][]{Condon1992,Peel2011,Bendo2015,Bendo2016,Salak2017,Michiyama2020,Chen2023}, enable us to quantify and model free-free, synchrotron and dust emission among galaxies in general and provide an direct approach to measure SFRs accurately using millimetre emission from the ALMA data. 


\begin{figure}
  \centering
  \includegraphics[width=0.48\textwidth]{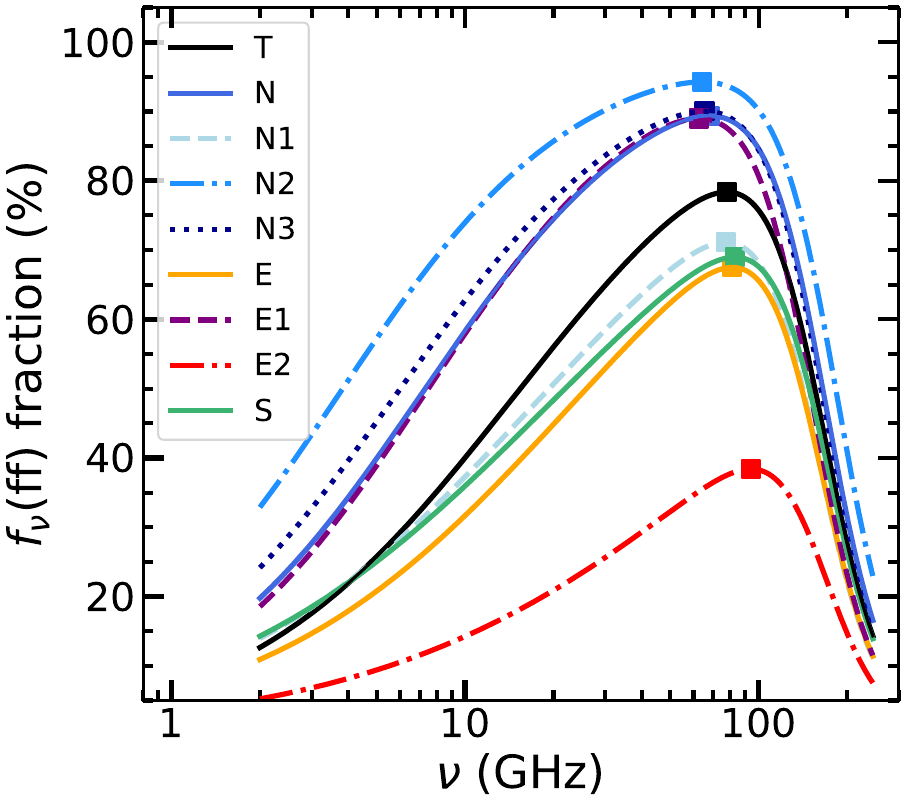} 
  \caption{The fraction of the continuum emission originating from free-free emission calculated within the different regions in the pseudobulge as a function of frequency. The squares show the frequencies where the fraction of emission from free-free emission reaches maximum values.}
  \label{fig:ff_fraction}
\end{figure}

We can use our results from NGC~1365 to at least show the best frequencies at which to detect free-free emission.  The variations in the fractions of continuum emission originating from free-free emission with frequencies for different regions of NGC~1365 are shown in Figure \ref{fig:ff_fraction}.  Notably, the maximum fractions of the continuum emission from free-free emission fall within the range of 60--100~GHz, which lies within the frequency range of ALMA Band 2 and 3 data.  
In addition, these maximum fractions are only marginally higher than the fractions measured at 103~GHz. For instance, in the T region, the maximum fraction is 78 per cent at 78~GHz while the fraction at 103~GHz is 75 per cent; the difference between these fractions is less than 5 per cent.  Hence, using the continuum emission in the 80--110~GHz range from ALMA Band 3 data for measuring the SFR of millimetre bright star-forming regions appears to be a viable option.

The fractions of the continuum emission originating from millimetre free-free emission typically range between 65 and 85 per cent, as observed in the N and S regions of NGC~1365, as well as the central starburst regions of NGC~253 \citep{Bendo2015}, NGC~4945 \citep{Bendo2016}, NGC~3256 \citep{Michiyama2020} and NGC~1808 \citep{Salak2017,Chen2023}.   Consequently, it is reasonable to assume the fraction of 80--100 GHz continuum emission from free-free sources is $75\pm 15$ per cent for typical star-forming regions ($\sim 0.5$--2~kpc).  This fraction may increase by about 15 per cent in an intensely starburst region, such as the N2 region in NGC~1365 and the southern starburst region in NGC~1808 \citep{Chen2023}.  Conversely, in regions where star formation is less significant, like the N1 and S regions, this fraction may decrease by about 10 per cent in the region, and it could decrease further if there exist significant outflows or radio jets, as observed in the E2 region, where this calibration method unsuitable for accurate SFR measurements.

As noted throughout our analysis, variations in the electron temperature could also affect the SFRs derived from millimetre free-free emission. 
Nonetheless, our work and the body of other work in the literature suggest that, assuming electron temperatures of $6000-10000$~K and free-free emission fractions of 75 per cent, it should be possible to measure extinction-free SFRs that are accurate to within 20 per cent.

\section{Conclusions} \label{sec:conclusion}

We have presented an analysis of the radio-to-submillimetre SEDs for the central pseudobulge of NGC~1365 based on archival ALMA and VLA data. 
Through the decomposition of the SED from 2 to 246~GHz, we demonstrate that free-free emission dominates of 50--120~GHz continuum emission from the pseudobulge, with synchrotron emission primarily observed at lower frequencies and thermal dust emission as the dominant emission source at higher frequencies.  In addition, no significant anomalous microwave emission or submillimetre excess emission is detected. 
The free-free emission from the photoionized gas primarily originates from the northern star-forming region as well as the fainter eastern and southwest star-forming regions. The thermal dust emission corresponds closely with the millimetre continuum emission at $\sim 100$ GHz with the exception of a dusty star-forming region in the northeastern part of the pseudobulge.
The synchrotron radiation is primarily associated with supernovae in the circumnuclear ring, although synchrotron emission is also detected from the AGN.  An outflow to the east of the AGN produces notably high synchrotron emission that has also been observed in $\oiii$ emission \citep[e.g.,][]{Venturi2017} and X-ray emission \citep[e.g.,][]{Wang2009}.
Notably, the fraction of the millimetre continuum emission originating from free-free emission varies significantly among the different subregions within the central pseudobulge, particularly for the eastern region that contains the outflow producing strong synchrotron emission. 

Free-free emission also dominates at the 103~GHz continuum emission in the central 380 pc region, but the photoionized gas associated with this emission is associated with the AGN rather than star formation.  We calculated that the observed free-free flux density is consistent with the broad Pa$\alpha$ line emission presented by \citet{Fazeli2019}.  However, the millimetre dust emission from the area around the AGN is relatively low because the dust is relatively hot and concentrated within a $\sim 50$ pc region \citep{Combes2019}.

The SFR within the central pseudobulge derived from the 103~GHz free-free emission is $8.9 \pm1.1$~M$_\odot$~yr$^{-1}$, and the star formation surface density averaged over this area is $1.6\pm 0.2$~M$_\odot$ yr$^{-1}$kpc$^{-2}$. The most intense star formation is observed in two star-forming regions in the northern part of the pseudobulge, where the star formation surface densities reach $12$~M$_\odot$ yr$^{-1}$kpc$^{-2}$.  

We treated the SFRs measured from the 103~GHz free-free emission as the best SFR measurements possible and compared them to other star formation metrics to test the effectiveness of the other metrics and understand how other physical processes could affect the accuracy of these metrics. 
The SFRs measured from the ALMA data are consistent with the SFRs from the mid-infrared data but significantly higher than the SFR derived from the extinction-corrected $\ha$ line emission.  This is likely because the central pseudobulge is so heavily dust obscured, so applying extinction corrections to the line emission is problematic.  The SFR averaged over the central pseudobulge estimated from the 1.5~GHz emission matches the SFR from the free-free emission, but the 1.5~GHz emission in the much smaller northern star-forming region yields a SFR much lower than the SFR measured from the free-free emission.  This difference is likely because synchrotron radiation produced by cosmic rays tend to be more extended than the star-forming regions \citep[e.g.,][]{Murphy2006}, while the central pseudobulge region is large enough to cover both the star-forming and more diffuse regions emitting the synchrotron emission.
Conversely, the 1.5~GHz emission in another smaller eastern region yields a SFR much higher than the SFR derived from the free-free emission, but this is because a significant fraction of the 1.5~GHz synchrotron emission from this region comes from an outflow from the AGN. 

Free-free emission stands out as a valuable SFR metrics because of its ability to directly trace young stellar populations while remaining generally unaffected by dust attenuation.  Radio-to-submillimetre SED analyses like what we have presented here for NGC~1365 could enable us to quantitatively model free-free emission among other galaxies in general and provide a direct approach to measuring SFRs accurately using ALMA millimetre data. 
Based on both the results from our analysis on NGC~1365 and other published analyses of the SEDs of other nearby starbursts, we propose that continuum emission from the 80--110~GHz range (which falls within ALMA Bands 2 and 3) multiplied by a 75 per cent correction factor could yield useful measurements of the free-free emission without performing any SED decomposition.  The SFRs based on these measurements would be accurate to within 20 per cent.  Several CMB survey telescopes, (e.g., the South Pole Telescope, the Atacama Cosmology Telescope, the Simons Observatory), the Green Bank Telescope, the Sardinia Radio Telescope, and the upcoming Square Kilometre Array (SKA) and Next Generation Very Large Array will provide abundant data for constructing more radio-to-submillimetre SEDs. In addition, ALMA Band 1 and future Band 2 observations, which will extend the frequency coverage of ALMA to 35~GHz, would place additional constraints on the SEDs measured for these galaxies and thus provide more accurate measurements of the free-free emission from extragalactic objects.

\section*{Acknowledgements}


We thank the reviewer and Dr. Patrick J. Leahy for their very useful comments on this paper.
GC acknowledges support from the SKA exchange programme of STFC and CSC (No. 201906340241).  
GJB is supported by STFC grant ST/T001488/1.
GAF acknowledges support from the Collaborative Research Centre 956, funded by the Deutsche Forschungsgemeinschaft (DFG) project ID 184018867.
This work is also supported by the National Science Foundation of China (NSFC, Grant No. 12233008, 11973038 and 11973039), the National Key R\&D Program of China (2023YFA1608100), the Strategic Priority Research Program of Chinese Academy of Sciences (Grant No. XDB 41000000), the China Manned Space Project (No. CMS-CSST-2021-A07, CMS-CSST-2021-B02), the Cyrus Chun Ying Tang Foundations, the 111 Project for "Observational and Theoretical Research on Dark Matter and Dark Energy" (B23042), and the CAS Pioneer Hundred Talents Program.

This paper makes use of the following ALMA data: ADS/JAO.ALMA\#2013.1.01161.S, 2015.1.01135.S, 2017.1.00129.S, 2019.1.01635.S, 2021.1.01150.S, 2021.1.01295.S, and 2021.2.00079.S. ALMA is a partnership of ESO (representing its member states), NSF (USA) and NINS (Japan), together with NRC (Canada), MOST and ASIAA (Taiwan), and KASI (Republic of Korea), in cooperation with the Republic of Chile.  The Joint ALMA Observatory is operated by ESO, AUI/NRAO and NAOJ.
The Australia Telescope Compact Array is part of the Australia Telescope National Facility (grid.421683.a) which is funded by the Australian Government for operation as a National Facility managed by CSIRO.

\section*{Data Availability}


The reduced, calibrated and science-ready ALMA data is available from the ALMA Science Archive at \url{https://almascience.eso.org/alma-data/archive}, and the science-ready VLA data is available from the National Radio Astronomy Observatory Data Archive at \url{https://data.nrao.edu} .



\bibliographystyle{mnras}
\bibliography{NGC1365} 

\begin{thebibliography}{}
\makeatletter
\relax
\def\mn@urlcharsother{\let\do\@makeother \do\$\do\&\do\#\do\^\do\_\do\%\do\~}
\def\mn@doi{\begingroup\mn@urlcharsother \@ifnextchar [ {\mn@doi@} {\mn@doi@[]}}
\def\mn@doi@[#1]#2{\def\@tempa{#1}\ifx\@tempa\@empty \href {http://dx.doi.org/#2} {doi:#2}\else \href {http://dx.doi.org/#2} {#1}\fi \endgroup}
\def\mn@eprint#1#2{\mn@eprint@#1:#2::\@nil}
\def\mn@eprint@arXiv#1{\href {http://arxiv.org/abs/#1} {{\tt arXiv:#1}}}
\def\mn@eprint@dblp#1{\href {http://dblp.uni-trier.de/rec/bibtex/#1.xml} {dblp:#1}}
\def\mn@eprint@#1:#2:#3:#4\@nil{\def\@tempa {#1}\def\@tempb {#2}\def\@tempc {#3}\ifx \@tempc \@empty \let \@tempc \@tempb \let \@tempb \@tempa \fi \ifx \@tempb \@empty \def\@tempb {arXiv}\fi \@ifundefined {mn@eprint@\@tempb}{\@tempb:\@tempc}{\expandafter \expandafter \csname mn@eprint@\@tempb\endcsname \expandafter{\@tempc}}}

\bibitem[\protect\citeauthoryear{{Ali-Ha{\"\i}moud}, {Hirata}  \& {Dickinson}}{{Ali-Ha{\"\i}moud} et~al.}{2009}]{Ali-Haimoud2009}
{Ali-Ha{\"\i}moud} Y.,  {Hirata} C.~M.,   {Dickinson} C.,  2009, \mn@doi [\mnras] {10.1111/j.1365-2966.2009.14599.x}, \href {https://ui.adsabs.harvard.edu/abs/2009MNRAS.395.1055A} {395, 1055}

\bibitem[\protect\citeauthoryear{{Anand} et~al.,}{{Anand} et~al.}{2021}]{Anand2021a}
{Anand} G.~S.,  et~al., 2021, \mn@doi [\mnras] {10.1093/mnras/staa3668}, \href {https://ui.adsabs.harvard.edu/abs/2021MNRAS.501.3621A} {501, 3621}

\bibitem[\protect\citeauthoryear{{Armus} et~al.,}{{Armus} et~al.}{2009}]{Armus2009}
{Armus} L.,  et~al., 2009, \mn@doi [\pasp] {10.1086/600092}, \href {https://ui.adsabs.harvard.edu/abs/2009PASP..121..559A} {121, 559}

\bibitem[\protect\citeauthoryear{{Battistelli} et~al.,}{{Battistelli} et~al.}{2019}]{Battistelli2019}
{Battistelli} E.~S.,  et~al., 2019, \mn@doi [\apjl] {10.3847/2041-8213/ab21de}, \href {https://ui.adsabs.harvard.edu/abs/2019ApJ...877L..31B} {877, L31}

\bibitem[\protect\citeauthoryear{{Bendo} et~al.,}{{Bendo} et~al.}{2006}]{Bendo2006}
{Bendo} G.~J.,  et~al., 2006, \mn@doi [\apj] {10.1086/508057}, \href {https://ui.adsabs.harvard.edu/abs/2006ApJ...652..283B} {652, 283}

\bibitem[\protect\citeauthoryear{{Bendo}, {Galliano}  \& {Madden}}{{Bendo} et~al.}{2012}]{Bendo2012}
{Bendo} G.~J.,  {Galliano} F.,   {Madden} S.~C.,  2012, \mn@doi [\mnras] {10.1111/j.1365-2966.2012.20784.x}, \href {https://ui.adsabs.harvard.edu/abs/2012MNRAS.423..197B} {423, 197}

\bibitem[\protect\citeauthoryear{{Bendo}, {Beswick}, {D'Cruze}, {Dickinson}, {Fuller}  \& {Muxlow}}{{Bendo} et~al.}{2015}]{Bendo2015}
{Bendo} G.~J.,  {Beswick} R.~J.,  {D'Cruze} M.~J.,  {Dickinson} C.,  {Fuller} G.~A.,   {Muxlow} T.~W.~B.,  2015, \mn@doi [\mnras] {10.1093/mnrasl/slv053}, \href {https://ui.adsabs.harvard.edu/abs/2015MNRAS.450L..80B} {450, L80}

\bibitem[\protect\citeauthoryear{{Bendo}, {Henkel}, {D'Cruze}, {Dickinson}, {Fuller}  \& {Karim}}{{Bendo} et~al.}{2016}]{Bendo2016}
{Bendo} G.~J.,  {Henkel} C.,  {D'Cruze} M.~J.,  {Dickinson} C.,  {Fuller} G.~A.,   {Karim} A.,  2016, \mn@doi [\mnras] {10.1093/mnras/stw1659}, \href {https://ui.adsabs.harvard.edu/abs/2016MNRAS.463..252B} {463, 252}

\bibitem[\protect\citeauthoryear{{Bendo}, {Miura}, {Espada}, {Nakanishi}, {Beswick}, {D'Cruze}, {Dickinson}  \& {Fuller}}{{Bendo} et~al.}{2017}]{Bendo2017}
{Bendo} G.~J.,  {Miura} R.~E.,  {Espada} D.,  {Nakanishi} K.,  {Beswick} R.~J.,  {D'Cruze} M.~J.,  {Dickinson} C.,   {Fuller} G.~A.,  2017, \mn@doi [\mnras] {10.1093/mnras/stx1837}, \href {https://ui.adsabs.harvard.edu/abs/2017MNRAS.472.1239B} {472, 1239}

\bibitem[\protect\citeauthoryear{{Bendo}, {Lu}  \& {Zijlstra}}{{Bendo} et~al.}{2020}]{Bendo2020}
{Bendo} G.~J.,  {Lu} N.,   {Zijlstra} A.,  2020, \mn@doi [\mnras] {10.1093/mnras/staa1589}, \href {https://ui.adsabs.harvard.edu/abs/2020MNRAS.496.1393B} {496, 1393}

\bibitem[\protect\citeauthoryear{{Bresolin}, {Schaerer}, {Gonz{\'a}lez Delgado}  \& {Stasi{\'n}ska}}{{Bresolin} et~al.}{2005}]{Bresolin2005}
{Bresolin} F.,  {Schaerer} D.,  {Gonz{\'a}lez Delgado} R.~M.,   {Stasi{\'n}ska} G.,  2005, \mn@doi [\aap] {10.1051/0004-6361:20053369}, \href {https://ui.adsabs.harvard.edu/abs/2005A&A...441..981B} {441, 981}

\bibitem[\protect\citeauthoryear{{CASA Team} et~al.,}{{CASA Team} et~al.}{2022}]{CASA2022}
{CASA Team} et~al., 2022, \mn@doi [\pasp] {10.1088/1538-3873/ac9642}, \href {https://ui.adsabs.harvard.edu/abs/2022PASP..134k4501C} {134, 114501}

\bibitem[\protect\citeauthoryear{{Calzetti} et~al.,}{{Calzetti} et~al.}{2010}]{Calzetti2010}
{Calzetti} D.,  et~al., 2010, \mn@doi [\apj] {10.1088/0004-637X/714/2/1256}, \href {https://ui.adsabs.harvard.edu/abs/2010ApJ...714.1256C} {714, 1256}

\bibitem[\protect\citeauthoryear{{Cepeda-Arroita} et~al.,}{{Cepeda-Arroita} et~al.}{2021}]{Cepeda-Arroita2021}
{Cepeda-Arroita} R.,  et~al., 2021, \mn@doi [\mnras] {10.1093/mnras/stab583}, \href {https://ui.adsabs.harvard.edu/abs/2021MNRAS.503.2927C} {503, 2927}

\bibitem[\protect\citeauthoryear{{Chang} et~al.,}{{Chang} et~al.}{2020}]{Chang2020}
{Chang} Z.,  et~al., 2020, \mn@doi [\apj] {10.3847/1538-4357/aba52f}, \href {https://ui.adsabs.harvard.edu/abs/2020ApJ...900...53C} {900, 53}

\bibitem[\protect\citeauthoryear{{Chen}, {Bendo}, {Fuller}, {Henkel}  \& {Kong}}{{Chen} et~al.}{2023}]{Chen2023}
{Chen} G.,  {Bendo} G.~J.,  {Fuller} G.~A.,  {Henkel} C.,   {Kong} X.,  2023, \mn@doi [\mnras] {10.1093/mnras/stad2450}, \href {https://ui.adsabs.harvard.edu/abs/2023MNRAS.525.3645C} {525, 3645}

\bibitem[\protect\citeauthoryear{{Combes} et~al.,}{{Combes} et~al.}{2019}]{Combes2019}
{Combes} F.,  et~al., 2019, \mn@doi [\aap] {10.1051/0004-6361/201834560}, \href {https://ui.adsabs.harvard.edu/abs/2019A&A...623A..79C} {623, A79}

\bibitem[\protect\citeauthoryear{{Condon}}{{Condon}}{1992}]{Condon1992}
{Condon} J.~J.,  1992, \mn@doi [\araa] {10.1146/annurev.aa.30.090192.003043}, \href {https://ui.adsabs.harvard.edu/abs/1992ARA&A..30..575C} {30, 575}

\bibitem[\protect\citeauthoryear{{Condon}, {Cotton}, {Greisen}, {Yin}, {Perley}, {Taylor}  \& {Broderick}}{{Condon} et~al.}{1998}]{Condon1998}
{Condon} J.~J.,  {Cotton} W.~D.,  {Greisen} E.~W.,  {Yin} Q.~F.,  {Perley} R.~A.,  {Taylor} G.~B.,   {Broderick} J.~J.,  1998, \mn@doi [\aj] {10.1086/300337}, \href {https://ui.adsabs.harvard.edu/abs/1998AJ....115.1693C} {115, 1693}

\bibitem[\protect\citeauthoryear{Cortes et~al.,}{Cortes et~al.}{2023}]{Cortes2023}
Cortes P.,  et~al., 2023, ALMA Cycle 10 Technical Handbook

\bibitem[\protect\citeauthoryear{{Crossley}, {Sjouwerman}, {Fomalont}  \& {Radziwill}}{{Crossley} et~al.}{2007}]{Crossley2007}
{Crossley} J.~H.,  {Sjouwerman} L.~O.,  {Fomalont} E.~B.,   {Radziwill} N.~M.,  2007, in American Astronomical Society Meeting Abstracts. p. 132.03

\bibitem[\protect\citeauthoryear{{Crossley}, {Sjouwerman}, {Fomalont}  \& {Radziwill}}{{Crossley} et~al.}{2008}]{Crossley2008}
{Crossley} J.~H.,  {Sjouwerman} L.~O.,  {Fomalont} E.~B.,   {Radziwill} N.~M.,  2008, in {Brissenden} R.~J.,  {Silva} D.~R.,  eds,  Society of Photo-Optical Instrumentation Engineers (SPIE) Conference Series Vol. 7016, Observatory Operations: Strategies, Processes, and Systems II. p. 70160O, \mn@doi{10.1117/12.787890}

\bibitem[\protect\citeauthoryear{{Davies}, {Dickinson}, {Banday}, {Jaffe}, {G{\'o}rski}  \& {Davis}}{{Davies} et~al.}{2006}]{Davies2006}
{Davies} R.~D.,  {Dickinson} C.,  {Banday} A.~J.,  {Jaffe} T.~R.,  {G{\'o}rski} K.~M.,   {Davis} R.~J.,  2006, \mn@doi [\mnras] {10.1111/j.1365-2966.2006.10572.x}, \href {https://ui.adsabs.harvard.edu/abs/2006MNRAS.370.1125D} {370, 1125}

\bibitem[\protect\citeauthoryear{{Dickinson} et~al.,}{{Dickinson} et~al.}{2018}]{Dickinson2018}
{Dickinson} C.,  et~al., 2018, \mn@doi [\nar] {10.1016/j.newar.2018.02.001}, \href {https://ui.adsabs.harvard.edu/abs/2018NewAR..80....1D} {80, 1}

\bibitem[\protect\citeauthoryear{{Draine}}{{Draine}}{2011}]{Draine2011}
{Draine} B.~T.,  2011, {Physics of the Interstellar and Intergalactic Medium}

\bibitem[\protect\citeauthoryear{{Draine} \& {Lazarian}}{{Draine} \& {Lazarian}}{1998}]{Draine1998}
{Draine} B.~T.,  {Lazarian} A.,  1998, \mn@doi [\apj] {10.1086/306387}, \href {https://ui.adsabs.harvard.edu/abs/1998ApJ...508..157D} {508, 157}

\bibitem[\protect\citeauthoryear{{Engelbracht} et~al.,}{{Engelbracht} et~al.}{2007}]{Engelbracht2007}
{Engelbracht} C.~W.,  et~al., 2007, \mn@doi [\pasp] {10.1086/521881}, \href {https://ui.adsabs.harvard.edu/abs/2007PASP..119..994E} {119, 994}

\bibitem[\protect\citeauthoryear{{Engelbracht}, {Rieke}, {Gordon}, {Smith}, {Werner}, {Moustakas}, {Willmer}  \& {Vanzi}}{{Engelbracht} et~al.}{2008}]{Engelbracht2008}
{Engelbracht} C.~W.,  {Rieke} G.~H.,  {Gordon} K.~D.,  {Smith} J. D.~T.,  {Werner} M.~W.,  {Moustakas} J.,  {Willmer} C.~N.~A.,   {Vanzi} L.,  2008, \mn@doi [\apj] {10.1086/529513}, \href {https://ui.adsabs.harvard.edu/abs/2008ApJ...678..804E} {678, 804}

\bibitem[\protect\citeauthoryear{{Fazeli}, {Busch}, {Valencia-S.}, {Eckart}, {Zaja{\v{c}}ek}, {Combes}  \& {Garc{\'\i}a-Burillo}}{{Fazeli} et~al.}{2019}]{Fazeli2019}
{Fazeli} N.,  {Busch} G.,  {Valencia-S.} M.,  {Eckart} A.,  {Zaja{\v{c}}ek} M.,  {Combes} F.,   {Garc{\'\i}a-Burillo} S.,  2019, \mn@doi [\aap] {10.1051/0004-6361/201834255}, \href {https://ui.adsabs.harvard.edu/abs/2019A&A...622A.128F} {622, A128}

\bibitem[\protect\citeauthoryear{{Fern{\'a}ndez-Torreiro} et~al.,}{{Fern{\'a}ndez-Torreiro} et~al.}{2023}]{Fernandez2023}
{Fern{\'a}ndez-Torreiro} M.,  et~al., 2023, \mn@doi [\mnras] {10.1093/mnras/stad2545}, \href {https://ui.adsabs.harvard.edu/abs/2023MNRAS.526.1343F} {526, 1343}

\bibitem[\protect\citeauthoryear{{Fern{\'a}ndez-Torreiro} et~al.,}{{Fern{\'a}ndez-Torreiro} et~al.}{2024}]{Fernandez2024}
{Fern{\'a}ndez-Torreiro} M.,  et~al., 2024, \mn@doi [\mnras] {10.1093/mnras/stad3145}, \href {https://ui.adsabs.harvard.edu/abs/2024MNRAS.527.11945} {527, 11945}

\bibitem[\protect\citeauthoryear{{Galametz} et~al.,}{{Galametz} et~al.}{2009}]{Galametz2009}
{Galametz} M.,  et~al., 2009, \mn@doi [\aap] {10.1051/0004-6361/200912963}, \href {https://ui.adsabs.harvard.edu/abs/2009A&A...508..645G} {508, 645}

\bibitem[\protect\citeauthoryear{{Galliano}, {Madden}, {Jones}, {Wilson}, {Bernard}  \& {Le Peintre}}{{Galliano} et~al.}{2003}]{Galliano2003}
{Galliano} F.,  {Madden} S.~C.,  {Jones} A.~P.,  {Wilson} C.~D.,  {Bernard} J.~P.,   {Le Peintre} F.,  2003, \mn@doi [\aap] {10.1051/0004-6361:20030814}, \href {https://ui.adsabs.harvard.edu/abs/2003A&A...407..159G} {407, 159}

\bibitem[\protect\citeauthoryear{{Galliano}, {Madden}, {Jones}, {Wilson}  \& {Bernard}}{{Galliano} et~al.}{2005}]{Galliano2005}
{Galliano} F.,  {Madden} S.~C.,  {Jones} A.~P.,  {Wilson} C.~D.,   {Bernard} J.~P.,  2005, \mn@doi [\aap] {10.1051/0004-6361:20042369}, \href {https://ui.adsabs.harvard.edu/abs/2005A&A...434..867G} {434, 867}

\bibitem[\protect\citeauthoryear{{Galliano} et~al.,}{{Galliano} et~al.}{2011}]{Galliano2011}
{Galliano} F.,  et~al., 2011, \mn@doi [\aap] {10.1051/0004-6361/201117952}, \href {https://ui.adsabs.harvard.edu/abs/2011A&A...536A..88G} {536, A88}

\bibitem[\protect\citeauthoryear{{Gao}, {Egusa}, {Liu}, {Kohno}, {Bao}, {Morokuma-Matsui}, {Kong}  \& {Chen}}{{Gao} et~al.}{2021}]{Gao2021}
{Gao} Y.,  {Egusa} F.,  {Liu} G.,  {Kohno} K.,  {Bao} M.,  {Morokuma-Matsui} K.,  {Kong} X.,   {Chen} X.,  2021, \mn@doi [\apj] {10.3847/1538-4357/abf738}, \href {https://ui.adsabs.harvard.edu/abs/2021ApJ...913..139G} {913, 139}

\bibitem[\protect\citeauthoryear{{Gordon} et~al.,}{{Gordon} et~al.}{2005}]{Gordon2005}
{Gordon} K.~D.,  et~al., 2005, \mn@doi [\pasp] {10.1086/429309}, \href {https://ui.adsabs.harvard.edu/abs/2005PASP..117..503G} {117, 503}

\bibitem[\protect\citeauthoryear{{Harper} et~al.,}{{Harper} et~al.}{2023}]{Harper2023}
{Harper} S.~E.,  et~al., 2023, \mn@doi [\mnras] {10.1093/mnras/stad1539}, \href {https://ui.adsabs.harvard.edu/abs/2023MNRAS.523.3471H} {523, 3471}

\bibitem[\protect\citeauthoryear{{Hensley}, {Murphy}  \& {Staguhn}}{{Hensley} et~al.}{2015}]{Hensley2015}
{Hensley} B.,  {Murphy} E.,   {Staguhn} J.,  2015, \mn@doi [\mnras] {10.1093/mnras/stv287}, \href {https://ui.adsabs.harvard.edu/abs/2015MNRAS.449..809H} {449, 809}

\bibitem[\protect\citeauthoryear{{Hermelo} et~al.,}{{Hermelo} et~al.}{2016}]{Hermelo2016}
{Hermelo} I.,  et~al., 2016, \mn@doi [\aap] {10.1051/0004-6361/201525816}, \href {https://ui.adsabs.harvard.edu/abs/2016A&A...590A..56H} {590, A56}

\bibitem[\protect\citeauthoryear{{Hirashita} \& {Ichikawa}}{{Hirashita} \& {Ichikawa}}{2009}]{Hirashita2009}
{Hirashita} H.,  {Ichikawa} T.~T.,  2009, \mn@doi [\mnras] {10.1111/j.1365-2966.2009.14726.x}, \href {https://ui.adsabs.harvard.edu/abs/2009MNRAS.396..500H} {396, 500}

\bibitem[\protect\citeauthoryear{{Hjelm} \& {Lindblad}}{{Hjelm} \& {Lindblad}}{1996}]{Hjelm1996}
{Hjelm} M.,  {Lindblad} P.~O.,  1996, \aap, \href {https://ui.adsabs.harvard.edu/abs/1996A&A...305..727H} {305, 727}

\bibitem[\protect\citeauthoryear{{Hunt}, {Bianchi}  \& {Maiolino}}{{Hunt} et~al.}{2005}]{Hunt2005}
{Hunt} L.,  {Bianchi} S.,   {Maiolino} R.,  2005, \mn@doi [\aap] {10.1051/0004-6361:20042157}, \href {https://ui.adsabs.harvard.edu/abs/2005A&A...434..849H} {434, 849}

\bibitem[\protect\citeauthoryear{{Kennicutt} \& {Evans}}{{Kennicutt} \& {Evans}}{2012}]{Kennicutt2012}
{Kennicutt} R.~C.,  {Evans} N.~J.,  2012, \mn@doi [\araa] {10.1146/annurev-astro-081811-125610}, \href {https://ui.adsabs.harvard.edu/abs/2012ARA&A..50..531K} {50, 531}

\bibitem[\protect\citeauthoryear{{Kroupa}}{{Kroupa}}{2002}]{Kroupa2002}
{Kroupa} P.,  2002, \mn@doi [Science] {10.1126/science.1067524}, \href {https://ui.adsabs.harvard.edu/abs/2002Sci...295...82K} {295, 82}

\bibitem[\protect\citeauthoryear{{Lee} et~al.,}{{Lee} et~al.}{2023}]{Lee2023}
{Lee} J.~C.,  et~al., 2023, \mn@doi [\apjl] {10.3847/2041-8213/acaaae}, \href {https://ui.adsabs.harvard.edu/abs/2023ApJ...944L..17L} {944, L17}

\bibitem[\protect\citeauthoryear{{Leitherer} et~al.,}{{Leitherer} et~al.}{1999}]{Leitherer1999}
{Leitherer} C.,  et~al., 1999, \mn@doi [\apjs] {10.1086/313233}, \href {https://ui.adsabs.harvard.edu/abs/1999ApJS..123....3L} {123, 3}

\bibitem[\protect\citeauthoryear{{Leitherer}, {Ekstr{\"o}m}, {Meynet}, {Schaerer}, {Agienko}  \& {Levesque}}{{Leitherer} et~al.}{2014}]{Leitherer2014}
{Leitherer} C.,  {Ekstr{\"o}m} S.,  {Meynet} G.,  {Schaerer} D.,  {Agienko} K.~B.,   {Levesque} E.~M.,  2014, \mn@doi [\apjs] {10.1088/0067-0049/212/1/14}, \href {https://ui.adsabs.harvard.edu/abs/2014ApJS..212...14L} {212, 14}

\bibitem[\protect\citeauthoryear{{Lindblad}}{{Lindblad}}{1999}]{Lindblad1999}
{Lindblad} P.~O.,  1999, \mn@doi [\aapr] {10.1007/s001590050018}, \href {https://ui.adsabs.harvard.edu/abs/1999A&ARv...9..221L} {9, 221}

\bibitem[\protect\citeauthoryear{{Linden} et~al.,}{{Linden} et~al.}{2019}]{Linden2019}
{Linden} S.~T.,  et~al., 2019, \mn@doi [\apj] {10.3847/1538-4357/ab2872}, \href {https://ui.adsabs.harvard.edu/abs/2019ApJ...881...70L} {881, 70}

\bibitem[\protect\citeauthoryear{{Linden}, {Murphy}, {Dong}, {Momjian}, {Kennicutt}, {Meier}, {Schinnerer}  \& {Turner}}{{Linden} et~al.}{2020}]{Linden2020}
{Linden} S.~T.,  {Murphy} E.~J.,  {Dong} D.,  {Momjian} E.,  {Kennicutt} R.~C. J.,  {Meier} D.~S.,  {Schinnerer} E.,   {Turner} J.~L.,  2020, \mn@doi [\apjs] {10.3847/1538-4365/ab8a4d}, \href {https://ui.adsabs.harvard.edu/abs/2020ApJS..248...25L} {248, 25}

\bibitem[\protect\citeauthoryear{{Liu} et~al.,}{{Liu} et~al.}{2023}]{Liu2023}
{Liu} D.,  et~al., 2023, \mn@doi [\apjl] {10.3847/2041-8213/aca973}, \href {https://ui.adsabs.harvard.edu/abs/2023ApJ...944L..19L} {944, L19}

\bibitem[\protect\citeauthoryear{{Maiolino} \& {Rieke}}{{Maiolino} \& {Rieke}}{1995}]{Maiolino1995}
{Maiolino} R.,  {Rieke} G.~H.,  1995, \mn@doi [\apj] {10.1086/176468}, \href {https://ui.adsabs.harvard.edu/abs/1995ApJ...454...95M} {454, 95}

\bibitem[\protect\citeauthoryear{{Michiyama}, {Iono}, {Nakanishi}, {Ueda}, {Saito}, {Yamashita}, {Bolatto}  \& {Yun}}{{Michiyama} et~al.}{2020}]{Michiyama2020}
{Michiyama} T.,  {Iono} D.,  {Nakanishi} K.,  {Ueda} J.,  {Saito} T.,  {Yamashita} T.,  {Bolatto} A.,   {Yun} M.,  2020, \mn@doi [\apj] {10.3847/1538-4357/ab88a5}, \href {https://ui.adsabs.harvard.edu/abs/2020ApJ...895...85M} {895, 85}

\bibitem[\protect\citeauthoryear{{Michiyama}, {Inoue}  \& {Doi}}{{Michiyama} et~al.}{2023}]{Michiyama2023}
{Michiyama} T.,  {Inoue} Y.,   {Doi} A.,  2023, \mn@doi [\pasj] {10.1093/pasj/psad044}, \href {https://ui.adsabs.harvard.edu/abs/2023PASJ...75..874M} {75, 874}

\bibitem[\protect\citeauthoryear{{Murphy} et~al.,}{{Murphy} et~al.}{2006}]{Murphy2006}
{Murphy} E.~J.,  et~al., 2006, \mn@doi [\apj] {10.1086/498636}, \href {https://ui.adsabs.harvard.edu/abs/2006ApJ...638..157M} {638, 157}

\bibitem[\protect\citeauthoryear{{Murphy} et~al.,}{{Murphy} et~al.}{2010}]{Murphy2010}
{Murphy} E.~J.,  et~al., 2010, \mn@doi [\apjl] {10.1088/2041-8205/709/2/L108}, \href {https://ui.adsabs.harvard.edu/abs/2010ApJ...709L.108M} {709, L108}

\bibitem[\protect\citeauthoryear{{Murphy} et~al.,}{{Murphy} et~al.}{2011}]{Murphy2011}
{Murphy} E.~J.,  et~al., 2011, \mn@doi [\apj] {10.1088/0004-637X/737/2/67}, \href {https://ui.adsabs.harvard.edu/abs/2011ApJ...737...67M} {737, 67}

\bibitem[\protect\citeauthoryear{{Murphy}, {Linden}, {Dong}, {Hensley}, {Momjian}, {Helou}  \& {Evans}}{{Murphy} et~al.}{2018}]{Murphy2018}
{Murphy} E.~J.,  {Linden} S.~T.,  {Dong} D.,  {Hensley} B.~S.,  {Momjian} E.,  {Helou} G.,   {Evans} A.~S.,  2018, \mn@doi [\apj] {10.3847/1538-4357/aac5f5}, \href {https://ui.adsabs.harvard.edu/abs/2018ApJ...862...20M} {862, 20}

\bibitem[\protect\citeauthoryear{Newville et~al.,}{Newville et~al.}{2022}]{Newville2022}
Newville M.,  et~al., 2022, lmfit/lmfit-py: 1.1.0, \mn@doi{10.5281/zenodo.7370358}

\bibitem[\protect\citeauthoryear{{Peel}, {Dickinson}, {Davies}, {Clements}  \& {Beswick}}{{Peel} et~al.}{2011}]{Peel2011}
{Peel} M.~W.,  {Dickinson} C.,  {Davies} R.~D.,  {Clements} D.~L.,   {Beswick} R.~J.,  2011, \mn@doi [\mnras] {10.1111/j.1745-3933.2011.01108.x}, \href {https://ui.adsabs.harvard.edu/abs/2011MNRAS.416L..99P} {416, L99}

\bibitem[\protect\citeauthoryear{{Pilbratt} et~al.,}{{Pilbratt} et~al.}{2010}]{Pilbratt2010}
{Pilbratt} G.~L.,  et~al., 2010, \mn@doi [\aap] {10.1051/0004-6361/201014759}, \href {https://ui.adsabs.harvard.edu/abs/2010A&A...518L...1P} {518, L1}

\bibitem[\protect\citeauthoryear{{Planck Collaboration} et~al.,}{{Planck Collaboration} et~al.}{2011}]{Planck2011_XVII}
{Planck Collaboration} et~al., 2011, \mn@doi [\aap] {10.1051/0004-6361/201116473}, \href {https://ui.adsabs.harvard.edu/abs/2011A&A...536A..17P} {536, A17}

\bibitem[\protect\citeauthoryear{{Planck Collaboration} et~al.,}{{Planck Collaboration} et~al.}{2015}]{Planck2015}
{Planck Collaboration} et~al., 2015, \mn@doi [\aap] {10.1051/0004-6361/201424643}, \href {https://ui.adsabs.harvard.edu/abs/2015A&A...582A..28P} {582, A28}

\bibitem[\protect\citeauthoryear{{Poidevin} et~al.,}{{Poidevin} et~al.}{2023}]{Poidevin2023}
{Poidevin} F.,  et~al., 2023, \mn@doi [\mnras] {10.1093/mnras/stac3151}, \href {https://ui.adsabs.harvard.edu/abs/2023MNRAS.519.3481P} {519, 3481}

\bibitem[\protect\citeauthoryear{{Poojon}, {Chung}, {Hoang}, {Baek}, {Nakanishi}, {Hirota}  \& {Tsai}}{{Poojon} et~al.}{2024}]{Poojon2024}
{Poojon} P.,  {Chung} A.,  {Hoang} T.,  {Baek} J.,  {Nakanishi} H.,  {Hirota} T.,   {Tsai} C.-W.,  2024, \mn@doi [\apj] {10.3847/1538-4357/ad1bc8}, \href {https://ui.adsabs.harvard.edu/abs/2024ApJ...963...88P} {963, 88}

\bibitem[\protect\citeauthoryear{{R{\'e}my-Ruyer} et~al.,}{{R{\'e}my-Ruyer} et~al.}{2013}]{Remy-Ruyer2013}
{R{\'e}my-Ruyer} A.,  et~al., 2013, \mn@doi [\aap] {10.1051/0004-6361/201321602}, \href {https://ui.adsabs.harvard.edu/abs/2013A&A...557A..95R} {557, A95}

\bibitem[\protect\citeauthoryear{{Rieke}, {Alonso-Herrero}, {Weiner}, {P{\'e}rez-Gonz{\'a}lez}, {Blaylock}, {Donley}  \& {Marcillac}}{{Rieke} et~al.}{2009}]{Rieke2009}
{Rieke} G.~H.,  {Alonso-Herrero} A.,  {Weiner} B.~J.,  {P{\'e}rez-Gonz{\'a}lez} P.~G.,  {Blaylock} M.,  {Donley} J.~L.,   {Marcillac} D.,  2009, \mn@doi [\apj] {10.1088/0004-637X/692/1/556}, \href {https://ui.adsabs.harvard.edu/abs/2009ApJ...692..556R} {692, 556}

\bibitem[\protect\citeauthoryear{{Risaliti}, {Elvis}, {Fabbiano}, {Baldi}  \& {Zezas}}{{Risaliti} et~al.}{2005}]{Risaliti2005}
{Risaliti} G.,  {Elvis} M.,  {Fabbiano} G.,  {Baldi} A.,   {Zezas} A.,  2005, \mn@doi [\apjl] {10.1086/430252}, \href {https://ui.adsabs.harvard.edu/abs/2005ApJ...623L..93R} {623, L93}

\bibitem[\protect\citeauthoryear{{Rosenberg}, {Wu}, {Le Floc'h}, {Charmandaris}, {Ashby}, {Houck}, {Salzer}  \& {Willner}}{{Rosenberg} et~al.}{2008}]{Rosenberg2008}
{Rosenberg} J.~L.,  {Wu} Y.,  {Le Floc'h} E.,  {Charmandaris} V.,  {Ashby} M.~L.~N.,  {Houck} J.~R.,  {Salzer} J.~J.,   {Willner} S.~P.,  2008, \mn@doi [\apj] {10.1086/524975}, \href {https://ui.adsabs.harvard.edu/abs/2008ApJ...674..814R} {674, 814}

\bibitem[\protect\citeauthoryear{{Sakamoto}, {Ho}, {Mao}, {Matsushita}  \& {Peck}}{{Sakamoto} et~al.}{2007}]{Sakamoto2007}
{Sakamoto} K.,  {Ho} P. T.~P.,  {Mao} R.-Q.,  {Matsushita} S.,   {Peck} A.~B.,  2007, \mn@doi [\apj] {10.1086/509775}, \href {https://ui.adsabs.harvard.edu/abs/2007ApJ...654..782S} {654, 782}

\bibitem[\protect\citeauthoryear{{Salak}, {Tomiyasu}, {Nakai}, {Kuno}, {Miyamoto}  \& {Kaneko}}{{Salak} et~al.}{2017}]{Salak2017}
{Salak} D.,  {Tomiyasu} Y.,  {Nakai} N.,  {Kuno} N.,  {Miyamoto} Y.,   {Kaneko} H.,  2017, \mn@doi [\apj] {10.3847/1538-4357/aa91cb}, \href {https://ui.adsabs.harvard.edu/abs/2017ApJ...849...90S} {849, 90}

\bibitem[\protect\citeauthoryear{{Sandqvist}, {Joersaeter}  \& {Lindblad}}{{Sandqvist} et~al.}{1995}]{Sandqvist1995}
{Sandqvist} A.,  {Joersaeter} S.,   {Lindblad} P.~O.,  1995, \aap, \href {https://ui.adsabs.harvard.edu/abs/1995A&A...295..585S} {295, 585}

\bibitem[\protect\citeauthoryear{{Schinnerer} et~al.,}{{Schinnerer} et~al.}{2023}]{Schinnerer2023}
{Schinnerer} E.,  et~al., 2023, \mn@doi [\apjl] {10.3847/2041-8213/acac9e}, \href {https://ui.adsabs.harvard.edu/abs/2023ApJ...944L..15S} {944, L15}

\bibitem[\protect\citeauthoryear{{Scoville} \& {Murchikova}}{{Scoville} \& {Murchikova}}{2013}]{Scoville2013}
{Scoville} N.,  {Murchikova} L.,  2013, \mn@doi [\apj] {10.1088/0004-637X/779/1/75}, \href {https://ui.adsabs.harvard.edu/abs/2013ApJ...779...75S} {779, 75}

\bibitem[\protect\citeauthoryear{{Silsbee}, {Ali-Ha{\"\i}moud}  \& {Hirata}}{{Silsbee} et~al.}{2011}]{Silsbee2011}
{Silsbee} K.,  {Ali-Ha{\"\i}moud} Y.,   {Hirata} C.~M.,  2011, \mn@doi [\mnras] {10.1111/j.1365-2966.2010.17882.x}, \href {https://ui.adsabs.harvard.edu/abs/2011MNRAS.411.2750S} {411, 2750}

\bibitem[\protect\citeauthoryear{{Stevens}, {Forbes}  \& {Norris}}{{Stevens} et~al.}{1999}]{Stevens1999}
{Stevens} I.~R.,  {Forbes} D.~A.,   {Norris} R.~P.,  1999, \mn@doi [\mnras] {10.1046/j.1365-8711.1999.02543.x}, \href {https://ui.adsabs.harvard.edu/abs/1999MNRAS.306..479S} {306, 479}

\bibitem[\protect\citeauthoryear{{Stevenson}}{{Stevenson}}{2014}]{Stevenson2014}
{Stevenson} M.~A.,  2014, \mn@doi [\apj] {10.1088/0004-637X/781/2/113}, \href {https://ui.adsabs.harvard.edu/abs/2014ApJ...781..113S} {781, 113}

\bibitem[\protect\citeauthoryear{{Storey} \& {Hummer}}{{Storey} \& {Hummer}}{1995}]{Storey1995}
{Storey} P.~J.,  {Hummer} D.~G.,  1995, \mn@doi [\mnras] {10.1093/mnras/272.1.41}, \href {https://ui.adsabs.harvard.edu/abs/1995MNRAS.272...41S} {272, 41}

\bibitem[\protect\citeauthoryear{{Swain}, {Shalima}  \& {Latha}}{{Swain} et~al.}{2023}]{Swain2023b}
{Swain} S.,  {Shalima} P.,   {Latha} K.~V.~P.,  2023, \mn@doi [\mnras] {10.1093/mnras/stad2770}, \href {https://ui.adsabs.harvard.edu/abs/2023MNRAS.tmp.2685S} {}

\bibitem[\protect\citeauthoryear{{Tabatabaei}, {Wei{\ss}}, {Combes}, {Henkel}, {Menten}, {Beck}, {Kov{\'a}cs}  \& {G{\"u}sten}}{{Tabatabaei} et~al.}{2013}]{Tabatabaei2013}
{Tabatabaei} F.~S.,  {Wei{\ss}} A.,  {Combes} F.,  {Henkel} C.,  {Menten} K.~M.,  {Beck} R.,  {Kov{\'a}cs} A.,   {G{\"u}sten} R.,  2013, \mn@doi [\aap] {10.1051/0004-6361/201321487}, \href {https://ui.adsabs.harvard.edu/abs/2013A&A...555A.128T} {555, A128}

\bibitem[\protect\citeauthoryear{{Turner}, {Urry}  \& {Mushotzky}}{{Turner} et~al.}{1993}]{Turner1993}
{Turner} T.~J.,  {Urry} C.~M.,   {Mushotzky} R.~F.,  1993, \mn@doi [\apj] {10.1086/173425}, \href {https://ui.adsabs.harvard.edu/abs/1993ApJ...418..653T} {418, 653}

\bibitem[\protect\citeauthoryear{{Venturi}, {Marconi}, {Mingozzi}, {Carniani}, {Cresci}, {Risaliti}  \& {Mannucci}}{{Venturi} et~al.}{2017}]{Venturi2017}
{Venturi} G.,  {Marconi} A.,  {Mingozzi} M.,  {Carniani} S.,  {Cresci} G.,  {Risaliti} G.,   {Mannucci} F.,  2017, \mn@doi [Frontiers in Astronomy and Space Sciences] {10.3389/fspas.2017.00046}, \href {https://ui.adsabs.harvard.edu/abs/2017FrASS...4...46V} {4, 46}

\bibitem[\protect\citeauthoryear{{Wang}, {Fabbiano}, {Elvis}, {Risaliti}, {Mazzarella}, {Howell}  \& {Lord}}{{Wang} et~al.}{2009}]{Wang2009}
{Wang} J.,  {Fabbiano} G.,  {Elvis} M.,  {Risaliti} G.,  {Mazzarella} J.~M.,  {Howell} J.~H.,   {Lord} S.,  2009, \mn@doi [\apj] {10.1088/0004-637X/694/2/718}, \href {https://ui.adsabs.harvard.edu/abs/2009ApJ...694..718W} {694, 718}

\bibitem[\protect\citeauthoryear{{Werner} et~al.,}{{Werner} et~al.}{2004}]{Werner2004}
{Werner} M.~W.,  et~al., 2004, \mn@doi [\apjs] {10.1086/422992}, \href {https://ui.adsabs.harvard.edu/abs/2004ApJS..154....1W} {154, 1}

\bibitem[\protect\citeauthoryear{{Whitmore} et~al.,}{{Whitmore} et~al.}{2023}]{Whitmore2023}
{Whitmore} B.~C.,  et~al., 2023, \mn@doi [\apjl] {10.3847/2041-8213/acae94}, \href {https://ui.adsabs.harvard.edu/abs/2023ApJ...944L..14W} {944, L14}

\makeatother
\end{thebibliography}




\appendix

\section{Spectra index of thermal dust emission}  \label{appendix:dust}


\begin{figure*}
  \centering
  \includegraphics[width=0.95\textwidth]{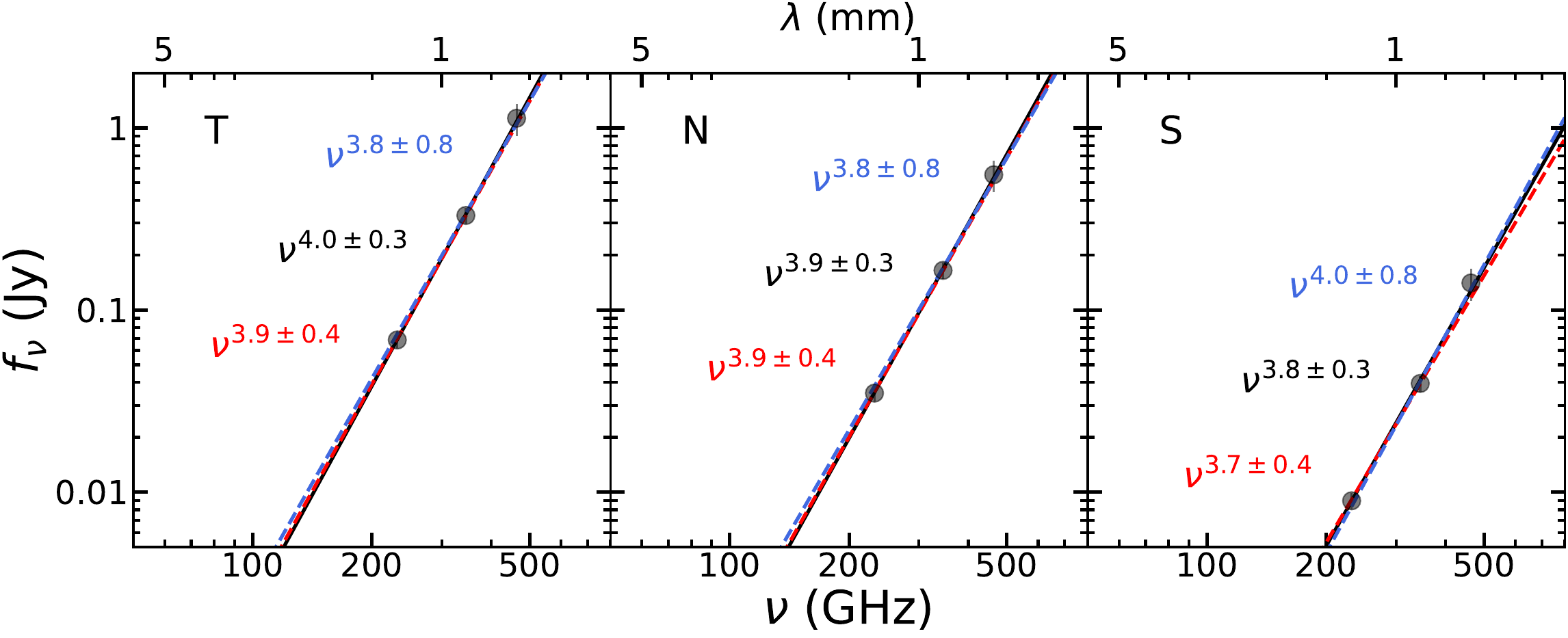}
  \caption{Continuum emission measurements across the ALMA Band 6, 7 and 8 for the T, N and S regions (from left to right). The best fits for the Band 6, 7 and 8 data are depicted with black solid lines. The red dashed lines represent the best fit based on the Band 6 and 7 data, while the blue dashed line represent the best fit from the Band 7 and 8 data. The best-fitting spectra indices, along with their $1\sigma$ uncertainties, are indicated as text that matches the colour of the corresponding best fit.}
  \label{fig:spectra-index_dust}
\end{figure*}

\begin{table*}
\centering
\begin{minipage}{0.725\textwidth}
\caption{ALMA observations and continuum measurements used in fitting the spectra index of dust emission.}
\label{tab:SED_dust}
    \begin{tabular}{@{}cccccccc@{}}
    \hline
    \hline
      Band &
      Rest-frame &
      Beam FWHM & 
      Beam FWHM & 
       & 
      Flux density 
       & \\
       &
      frequency &
      before smoothing & 
      after smoothing & 
      T region & 
      N region & 
      S region & 
        \\
       &
      (GHz) &
      ($\rm {arcsec^2}$) &
      ($\rm {arcsec^2}$) &
      (mJy) &
      (mJy) &
      (mJy) \\
    \hline  
      6 & 231.8 & 2.2 $\times$ 1.9 & 4.5 $\times$ 2.5 & 68 $\pm$ 7 & 35 $\pm$ 4 & 9 $\pm$ 1  \\  
      7 & 345.3 & 4.7 $\times$ 2.4 & - & 331 $\pm$ 33 & 165 $\pm$ 16 & 40 $\pm$ 4\\  
      8 & 463.6 & 3.2 $\times$ 2.3 & 4.5 $\times$ 2.6 & 1130 $\pm$ 226 & 553 $\pm$ 111 & 141 $\pm$ 28 \\  
    \hline
    \end{tabular}\\
\end{minipage}
\end{table*}

To constrain the spectral slope of the Rayleigh-Jeans side of the dust emission within NGC~1365 and to assess whether the slope may flatten, which would be indicative of ``submillimetre excess'' emission \citep[e.g.,][]{Galliano2003,Galliano2005,Bendo2006,Galametz2009,Galliano2011,Planck2011_XVII,Hermelo2016,Chang2020}, we performed additional analyses of the SED of NGC~1365 using the ALMA Band 6 data listed in Table~\ref{tab:ALMA_obs} as well as ALMA Band 7 from project 2021.2.00079.S and Band 8 data from project 2019.1.01635.S.
Compared to the Band 6 data, the Band 7 and 8 data have larger beam FWHMs and smaller maximum recoverable scales, which is why we did not incorporate those data into the analysis in Section~\ref{sec:SED}.  To create images of these data for this analysis, we first adjusted the $uv$ coverage of the data in each band to match the maximum recoverable scale from the Band 8 data before imaging the data.  Next, when imaging the Band 7 and 8 data using {\sc tclean}, we set the {\sc weighting} parameter to {\sc briggs} and the {\sc robust} parameter to 0.5.  After this, we used {\sc imsmooth} in {\sc casa} version 6.4.1 to smooth the beams from the Band 6 and 8 data to match the resolution in the Band 7 data.  The continuum measurements for the T, N and S regions used in this analysis are listed in Table \ref{tab:SED_dust}.

The continuum measurements were fitted as a single power law model in logarithmic space.  Figure \ref{fig:spectra-index_dust} show the best-fitting results for the T, N and S regions. The best fits for the Band 6, 7 and 8 data are indicated by black lines, while the best fit for the combined Band 6 and 7 data is shown in red, and the Band 7 and 8 data in blue.  These comparable spectral indices suggest that the continuum emission within this frequency range is well characterized by a single power law with a spectral index of $\approx$4.0.  This would be consistent with $\beta\geq2.0$, which is what is expected for standard dust emission.  If submillimetre excess emission was present, the spectral slope of these data would be expected to be shallower.  Additionally, we see no inflection in the slope of the data between ALMA Bands 6 and 8, which indicates that no significant submillimetre excess emission appears between these bands.

Based on the results from this analysis, we adopted a spectral index of 4.0 for the thermal dust emission in our SED fits in Section \ref{sec:SED}.  The uncertainty in the spectral slope is 0.3, but this will have a minimal effect on our main results.  When accounting for this uncertainty in the SED fits, the contribution of free-free emission to the 103 GHz continuum emission fluctuated by less than 5 per cent.

\section{Non detection of AME on the pseudobulge of NGC~1365}  \label{appendix:ame}

\begin{table*}
\centering
\begin{minipage}{0.625\textwidth}
\caption{General information of observations and continuum measurements for the T region.}
\label{tab:SED_AME}
    \begin{tabular}{@{}cccccc@{}}
    \hline
    \hline
      Instrument &
      Band &
      Rest-frame &
      Beam FWHM & 
      Beam FWHM & 
      Flux density \\
       &
       &
      frequency &
      before smoothing & 
      after smoothing & 
        \\
       &
       &
      (GHz) &
      ($\rm {arcsec^2}$) &
      ($\rm {arcsec^2}$) &
      (mJy) \\
    \hline  
      VLA 
      & S & 2.6 & 2.6 $\times$ 0.8 & 10.3 $\times$ 10.0 & 202 $\pm$ 20 \\
      &   & 3.6 & 2.2 $\times$ 0.8 & 10.2 $\times$ 10.0 & 156 $\pm$ 16 \\
      &   & 4.9 & 5.5 $\times$ 1.5 & 10.6 $\times$ 9.4 & 127 $\pm$ 13 \\   
      & C & 5.0 & 1.7 $\times$ 0.5 & 10.2 $\times$ 10.0 & 118 $\pm$ 12 \\   
      &   & 7.0 & 1.3 $\times$ 0.4 & 10.2 $\times$ 10.0 & 90 $\pm$ 9 \\   
      & X & 9.0 & 2.3 $\times$ 0.6 & 10.3 $\times$ 10.0 & 73 $\pm$ 7 \\   
      & Ku& 15.0 & 5.5 $\times$ 1.5 & 10.6 $\times$ 9.4 & 54 $\pm$ 5 \\ 
      & K & 22.6 & 11.1 $\times$ 1.2 & 13.7 $\times$ 8.3 & 33 $\pm$ 5 \\
    \hline
      ATCA 
      & C & 5.6 & 3.5 $\times$ 1.2 & 10.6 $\times$ 10.1 & 125 $\pm$ 13 \\
      & K & 23.4 & 6.1 $\times$ 1.2 & 10.9 $\times$ 9.1 & 36 $\pm$ 5 \\
    \hline
      ALMA 
      & 3 & 103.1 & 2.3 $\times$ 1.8 & 10.3 $\times$ 10.2 & 25.1 $\pm$ 1.3 \\ 
      &   & 114.1 & 2.3 $\times$ 1.8 & 10.3 $\times$ 10.2 & 27.7 $\pm$ 1.4 \\  
      & 6 & 231.8 & 2.2 $\times$ 1.9 & 10.2 $\times$ 10.1 & 93 $\pm$ 9 \\   
      &   & 246.3 & 2.2 $\times$ 1.9 & 10.2 $\times$ 10.2 & 114 $\pm$ 11 \\
    \hline
    \end{tabular}\\
\end{minipage}
\end{table*}


Anomalous microwave emission (AME) at 20 to 60~GHz has been detected from several nearby galaxies, including M31 \citep{Planck2015,Battistelli2019,Harper2023,Fernandez2024}, NGC~2903 \citep{Poojon2024}, NGC 4725 \citep{Murphy2018} and NGC 6946 \citep{Murphy2010,Hensley2015}.  To investigate the potential contribution of AME within the pseudobulge of NGC 1365 (the T region), we performed additional analyses of the SED using the ALMA Band 3 and 6 data; the VLA S-, C-, X-, Ku- and K-band data; and the ATCA C- and K-band data listed in Table~\ref{tab:SED_AME}.  The VLA Ku- and K-band data were processed using the the same steps described in Sections \ref{subsec:VLA}.  We followed the the ATCA Users Guide\footnote{\url{https://www.narrabri.atnf.csiro.au/observing/users_guide/html/}}, to calibrate the ATCA data using the {\sc miriad} package\footnote{\url{https://www.atnf.csiro.au/computing/software/miriad/}}. The calibrated data were then converted into continuum images using {\sc tclean} in {\sc casa} version 6.4.1, with the {\sc weighting} parameter set to {\sc briggs} and the {\sc robust} parameter set to 0.5 to more closely match the beam FWHM to the FWHM of the other data.
When creating all of the continuum images for this analysis, the $uv$ coverages were adjusted so that all images have the same maximum recoverable scale (36 arcsec).  After this, we used {\sc imsmooth} in {\sc casa} version 6.4.1 to smooth the beams from  all the data to match the resolution with that of the VLA K-band data.

The spinning dust model effectively describe the majority of AME \citep{Draine1998,Ali-Haimoud2009}, yet the specific mechanisms for this emission remains ambiguous.  AME has been widely characterized through an empirical lognormal approximation, which was first introduced by \citet{Stevenson2014} and later developed by \citet{Cepeda-Arroita2021}. 
This approach simplifies the {\sc SPDUST} model \citep{Ali-Haimoud2009,Silsbee2011} by introducing only three free parameters, thus avoiding degeneracies associated with the models.  To include AME within our SED models (as presented in Equation \ref{eq:SED}), the equation describing the overall SED is given as
\begin{equation}
\begin{split}
  f_\nu = {}&f_{\nu}(\mathrm{syn}) + f_\nu(\mathrm{ff}) + f_{\nu}(\mathrm{dust}) + f_{\nu}(\mathrm{AME}) \\
={}&A_{\mathrm{syn}} \left(\frac{\nu}{1\ \mathrm{GHz}}\right)^{\alpha_\mathrm{syn}}
+A_{\mathrm{ff}} g_{\mathrm{ff}} + A_{\mathrm{dust}} \left(\frac{\nu}{200\ \mathrm{GHz}}\right)^{\alpha_\mathrm{dust}} \\ 
{}&+ A_{\mathrm{AME}} \mathrm{exp}\left[-\frac{1}{2W_{\mathrm{AME}}^2} \ln^2\left(\frac{\nu}{\nu_{\mathrm{AME}}}\right) \right].
\end{split}
\label{eq:SED-AME}
\end{equation}
For the AME peak, the flux density ($A_{\rm AME}$) is set to be a non-negative parameter, the frequency ($\nu_{\rm AME}$) is constrained to a range of 10 to 60~GHz, and the width ($W_{\rm AME}$) is constrained to between 0.2 and 1.  These constraints, obtained from the comparisons with {\sc SPDUST} models, have been extensively applied in detecting AME \citep{Cepeda-Arroita2021,Poidevin2023,Fernandez2023,Fernandez2024,Poojon2024}. Following the methodology described in Section \ref{sec:SED}, we used the Levenberg-Marquardt algorithm with the {\sc LMFIT} package to fit the SED in logarithmic space, and then we sampled the posterior distributions of the fitting parameters.

\begin{figure*}
  \centering
  \includegraphics[width=1.0\textwidth]{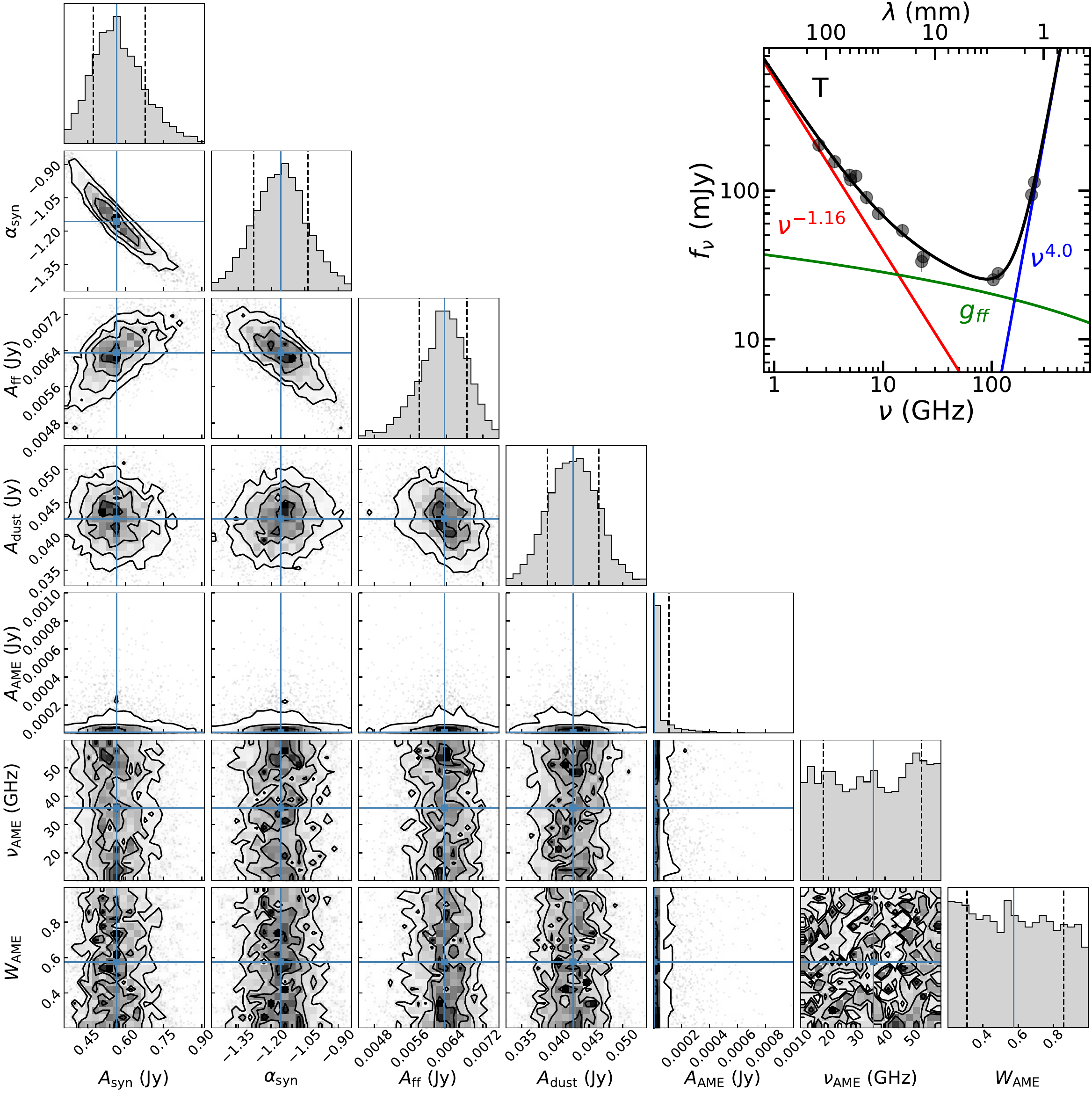}
  \caption{
  The SED (top-right) and the corresponding corner plot (left) for the T region of NGC~1365.  The SED is similar to the one shown in Figure ~\ref{fig:SED_T}, but it also includes VLA Ku- and K-band data and the ATCA C- and K-band data to assist in constraining the potential contribution of AME.  The corner plot shows the posterior distributions for parameters in Equation \ref{eq:SED-AME} characterizing the SED.}
  \label{fig:SED_fit_AME}
\end{figure*}

The SED, along with the best fitting functions and the corner plot for the marginalized posterior distributions and correlation diagrams are illustrated in Figure \ref{fig:SED_fit_AME}.  The SED is predominantly represented through a combination of synchrotron, free-free and thermal dust emission, and free-free emission still dominates between 30 and 150 GHz; $78 \pm 5$ per cent of the 103 GHz continuum emission is produced by the free-free emission.  However, no significant AME is detected; $A_{\rm AME}$ is $\approx 0$, and $\nu_{\rm AME}$ and $W_{\rm AME}$ are poorly constrained, as illustrated in the posterior distributions.  The fraction of the millimetre continuum emission originating from AME is less than 1 per cent.
Furthermore, the Bayesian information criterion (BIC) value for this SED model is 9.7, in contrast to a significantly lower BIC value of 1.7 when the same SED is fitted without including the AME component in the model.

Observations of AME usually correlate with the presence of abundant diffuse dust within nearby galaxies \citep[e.g.,][]{Dickinson2018}.  Additionally, these galaxies or the regions within these galaxies exhibit weak star formation.  In contrast, the pseudobulge of NGC 1365 contains a starburst that is notably associated with significant free-free emission.  This difference might explain why AME could be detected in galaxies like M31 but not in starburst galaxies such as M82 and NGC~253 \citep{Peel2011} as well as LIRGs such as Arp 220, Mrk 231, NGC 3690, NGC 6240 \citep{Dickinson2018} and NGC 1365 in this work.  Given the lack of data within the 30-80 GHz frequency range, the presence of AME within the pseudobulge of NGC 1365 cannot be entirely ruled out.  However, our SED analysis suggests that if AME is present, the AME would be different from those detected in other nearby galaxies; either the peak of AME would be at frequencies $\gtrsim$ 60 GHz or the width of the AME ($W_{\mathrm{AME}}$) is outside the range of 0.2 to 1.0 used in SED fits applied to other galaxies.  Consequently, given the negligible contribution of AME, we did not include the contribution of AME in the SED model presented in Section \ref{sec:SED}. 

\bsp	
\label{lastpage}
\end{document}